# Topologically Protected Vortex Structures to Realize Low-Noise Magnetic Sensors

Dieter Suess[1], Anton Bachleitner-Hofmann[1], Armin Satz[2], Herbert Weitensfelder[1], Christoph Vogler[1], Florian Bruckner[1], Claas Abert[1], Klemens Prügl[4], Jürgen Zimmer[3], Christian Huber[2], Sebastian Luber[3], Wolfgang Raberg[3], Thomas Schrefl[5], Hubert Brückl[5]

[1]University of Vienna, Christian Doppler Laboratory, Faculty of Physics, Physics of Functional Materials, Währinger Straße 17, 1090 Vienna, Austria

[2]Infineon Technologies AG, Siemensstraße 2, 9500 Villach, Austria

[3]Infineon Technologies AG, Am Campeon 1-12, 85579 Neubiberg, Germany

[4]Infineon Technologies AG, Wernerwerkstrasse 2, 93049 Regensburg, Germany

[5]Center for Integrated Sensor Systems, Danube University Krems, Viktor Kaplan Str. 2 E, 2700 Wiener Neustadt, Austria

***Abstract:*** Micromagnetic sensors play a major role towards the miniaturization in the industrial society. The adoption of new and emerging sensor technologies like anisotropic magnetoresistance (AMR), giant magnetoresistance (GMR) and tunnel magnetoresistance (TMR) sensors[1] are mainly driven by their integrability and enhanced sensitivity. At the core of such sensors, a microstructured ferromagnetic thin film element transduces the magnetic signal. Such elements usually switch via multi-domain, C- or S-shaped magnetization states and, therefore, often exhibit an open non-linear hysteresis curve. Linearity and hysteretic effects, as well as magnetic noise are key features in the improvement of such sensors. Here, we report on the physical origin of these disturbing factors and the inherent connection of noise and hysteresis. Critical noise sources are identified by means of analytic and micromagnetic models. The dominant noise source is due to irreproducible magnetic switching of the transducer element at external fields close to the Stoner–Wohlfarth switching field. Furthermore, a solution is presented to overcome these limiting factors: a disruptive sensor design is proposed and analyzed which realizes a topologically protected magnetic vortex state[2] in the transducer element. Compared to state of the art sensors the proposed sensor layout has negligible hysteresis, a linear regime about an order of magnitude higher and lower magnetic noise making the sensor ideal candidate for applications ranging from automotive industry[3] to biological application [4].

## 1. Introduction

In modern society, various applications rely on magnetic forces to move parts in electric vehicles or generators and to store data on hard disks. However, besides using magnetic fields as transmitters of mechanical forces, other important applications of magnets involve the detection of magnetic fields. The total market of semiconductor magnetic field sensors, which include Hall, giant magnetoresistance[5–7] (GMR), anisotropic magnetoresistance[8] (AMR), and tunnel magnetoresistance (TMR) sensors[1,7,9], is about 1670 Mio USD, of which the automotive market accounts for about 930 Mio USD [10]. Other main applications are wireless communication (~292 Mio USD), manufacturing & process automation, consumer appliances, and military & civil aerospace accounts. GMR sensors were suggested for position and speed detection in industrial and automotive applications [3,11] and are nowadays used for these applications worldwide. Still, the market for application of xMR (GMR & TMR) is increasing tremendously due to the small size, low power consumption, and CMOS compatibility. Nevertheless, there are significant open questions and challenges that have already been reported in the early works but that are still relevant nowadays. The most important properties of xMR magnetic sensors are: (i) a large linear sensing regime in order to relax design guidelines for the field-generating devices, such as permanent magnets, (ii) a small hysteresis, and (iii) low noise, leading to a good signal-to-noise ratio. State-of-the-art xMR sensors rely on the shape anisotropy of thin microstructured elliptical or rectangular elements, as schematically illustrated in *Fig. 1* (a). The easy axis of the free layer (blue) is parallel to the long axis of the elliptical particle due to shape anisotropy. The free layer and pinned layer (violet) are separated by a spacer layer. The pinned layer is magnetized orthogonally to the long axis. The magnetic state of the pinned layer is stabilized by antiferromagnets, such as PtMn or IrMn. In order to reduce the stray field of the pinned layer, synthetic antiferromagnets are typically used [12]. If the free layer can be described by an ideal Stoner–Wohlfarth (SW) particle with homogeneous magnetization, the transfer curve is perfectly linear until the anisotropy field $H_k$ is reached. A conventional xMR free layer, however, typically exhibit a hysteresis with a finite coercive field $H_c$ (Fig. 1 (a)). The saturation field is denoted by $H_s$. Experimentally, the best signal-to-noise ratios are obtained by using flux concentrators that lead to higher sensitivity but reduced the linear range[13,14].

In applications, magnetic sensors are exposed to complicated magnetic field profiles. In the following, we will discuss a typical field profile that occurs in automotive pole wheels for speed detection. It is an elliptical rotating external field within the plane of the sensor with frequency $\omega = 2\pi f$:

$$\mathbf{H}_{\text{ext}} = \begin{pmatrix} A_x \cdot \cos(\omega t) \\ A_y \cdot \sin(\omega t) \end{pmatrix} \tag{1}$$

Due to the large shape anisotropy of the several-nanometer-thick free layer, the *z*-component of a magnetic field can be neglected. If the amplitude $A_y = 0$, a collinear field is applied along the hard axis of the free layer. This describes the typical operation mode in a hard disk drive. A magnetic encoder wheel generates a field profile according to Eq. (1) with $A_y \neq 0$. The amplitudes $A_x$ and $A_y$ can vary over a wide range, depending on the relative position of the encoder wheel and magnetic sensor, individual after their montage. A typical mode of operation is to detect the zero crossing of the resistance $R$ of the sensor, where $R$ is normalized to +-1. It is important in the application that this zero crossing, caused by a periodic shift from the magnetic north to south pole due to the rotation of the encoder wheel, is as reproducible as possible. Since the slope of the $R(t)$ curve at $R = 0$ is positive and negative, we denote this zero crossing with a rising edge and falling edge, respectively. In this paper, all phase noise is evaluated at the falling edge. The standard deviation of the normalized period length, as defined in Eq. (8), is further referred to as phase noise. Since the frequency of the zero crossing is used to extract the rotation speed of the wheel, small phase noise is required for high accuracy.

The experimental phase noise as a function of the field amplitudes $A_x$ and $A_Y$ is shown in Fig. 2 (e1). Details of the phase noise measurement are given in the method section. The sensor stack consists of a synthetic antiferromagnetic (SA) reference system (pinned layer) of two CoFe layers that are antiparallel coupled via a Ru layer. The SA is pinned to the antiferromagnetic PtMn layer. A Cu layer is used as a spacer layer. The free layer consists of a CoFe-based alloy. The sensor stack is contacted by vias as shown schematically in Fig. 2 (v3) that are connected through the substrate to the reference system. The current flow is in plane the sensor stack (CIP). The phase noise is small for large field amplitudes $A_x$ and $A_y$ (point (C) in fig. 2). Interestingly, also for small field amplitudes, the phase noise is small (point (A)). However, a surprisingly large noise appears for intermediate field amplitudes(point (B)). For a particular region of $A_x$ and $A_y$, the phase noise is about 10 times higher compared to the phase noise floor. Due to the shape of the phase noise region in Fig. 2 (e1), we denote it in the following as phase noise croissant.

This phase noise croissant is the dominant noise source of state-of-the-art xMR sensors and the main limiting factor concerning the accuracy of wheel speed sensors.

In the following, we will elucidate the origin of this noise source and show that phase noise is a fundamental property of microstructured magnetic elements with a homogeneous magnetization exposed to external fields with a certain rotational field strength.

## 2. Phase noise due thermal fluctuations: Stoner Wohlfarth element

In order to simulate the switching behavior at finite temperature, we use a kinetic Monte Carlo approach, [15,16] which is explained in detail in the method section. This procedure can be used in order to calculate $M_x(t)$ for different field amplitudes $A_x$ and $A_y$. In Fig. 3 (a-c), the blue curves show the $M_x(H_x)$ response for three different field strengths of the rotating field calculated with this method. For the SW particle, an anisotropy field $H_k$ = 10mT is assumed, and the easy axis is tilted by $\zeta$ = 2° with respect to the easy axis. The energy barrier in the SW particle is assumed to be $\Delta E = 230 k_B T_{300}$ . Fig. 3 (a) shows the response for a rotating field that is entirely within the SW astroid. Hence, the magnetic state remains in the up state, and a reproducible response can be observed. For fields that intersect with the SW astroid, a reproducible zero crossing of $M_x(H_x)$ can also be observed, as shown in Fig. 3 (c). The situation at the phase noise croissant where the rotating field is close to the SW astroid is shown in Fig. 3 (b). Two different zero crossings of the $M_x(H_x)$ curve can be observed, corresponding to the situation when the particle is in the up or down state, respectively. Furthermore, the thermally activated switching process can clearly be seen by the irreversible jump in the blue curve at a different field strength $H_x$.

In order to calculate phase noise, the $M_x(t)$ is calculated for 100 rotations of the external field, and the zero crossings of $M_x(t)$ of the falling edge are evaluated. Using the zero crossings, the phase noise is calculated according to Eq. (8) for a frequency of 1 kHz. The obtained phase noise as a function of $A_x$ and $A_y$ is shown in Fig. 2 (e3). This procedure opens a corridor at $T$ > 0 K, where the thermally activated switching into the other state (from down to up or vice versa) may or may not happen. This contribution of randomness can be observed as phase noise. At very low temperatures, the corridor is very sharp and almost coincident with the calculated $T$=0 K switching relation. However, with rising temperature, the corridor increases in width and shifts to lower fields.

A detailed explanation, why phase noise happens in a SW particle is given in the supplementary material in the section "Fundamental phase noise in a Stoner-Wohlfarth model".

Interestingly, it can be seen that this model qualitatively shows high phase noise in regions close to the experiment. It demonstrates the important insight that phase noise is an inherent effect, even in single domain particles, and that for a finite angle between the $y$-axis and the easy axis, phase noise cannot be reduced to zero, even if the elements are perfectly fabricated without any structural defects. Furthermore, it can be seen that the phase noise splits into two branches. The reason is that for a constant $A_x$ amplitude, there are two different $A_y$ amplitudes that result in rotating fields, tangential to the SW astroid. The splitting becomes larger for rotating fields with higher aspect ratios and vanishes for circular rotating fields.

From experiments, as well as from simulations, it was shown that the phase noise of $N$ elements $\sigma_{\tau,N}$ is connected to the phase noise of one element $\sigma_{\tau,1}$, according to:

$$\sigma_{\tau,N} = \frac{\sigma_{\tau,1}}{\sqrt{N}} \quad (2)$$

It has to be noted that the calculated phase noise is evaluated for one elliptical particle; however, in the experiment, the number of particles is 1752. Hence, the corresponding phase noise for the same number of particles is significantly smaller in the simulation.

## 3. Phase noise due to domain processes: micromagnetic model

The previous discussion shows that for any single domain particle, phase noise inevitably occurs for particular ratios of the $A_x$ and $A_y$ amplitudes. However, the predicted values are significantly smaller than the experimental measurements. Hence, in the following, we extend the model, taking into account inhomogeneous magnetization during reversal using micromagnetics simulations using a finite difference code [17,18]. Since the rotation speed in the application, which is in the range of up to 10 kHz, is significantly smaller than the intrinsic micromagnetic precessional frequency, it is not necessary to resolve the dynamics in the simulations. The simulation speed can be drastically increased if, instead of the solution of the Landau–Lifshitz–Gilbert equation[19], only equilibrium states are calculated by minimizing the total energy by using a limited-memory quasi-Newton method with a backtracking line search with an extrapolated initial step[20].

The finite difference in mesh size in the simulation is 5 nm. The saturation magnetization of $J_s$ = 1.75 T was extracted from the VSM measurement. The exchange constant was assumed to be $A_{ex}$ = 15 pJ/m. No crystalline easy axis was assumed. The rotation of the external field is simulated by discretizing a full rotation into $N$ steps so that $\Delta H_x$ < 0.05 mT. For each of these $N$ field directions the energy is minimized until convergence was achieved, as defined in Ref[[20]]. 10 full rotations are simulated in order to extract the phase noise. An edge roughness is included in the simulations by discretizing the edge in segments of 5 nm and randomly displacing the respective segment joints along the edge normal, according to a normal distribution with a standard deviation of 5 nm. In Fig. 3 (a-c), the $M_x(H_x)$ response of the micromagnetic simulations is shown for three distinct cases: (a) fields smaller than the phase noise croissant, (b) fields at the phase noise croissant, and (c) fields larger than the phase noise croissant. For field cases (a) and (c), the expected response is obtained, which qualitatively agrees well with the SW model and also with the experimentally obtained data (green).

A very interesting effect can be observed for the intermediate field strength case (b). Although the simulations are performed at $T = 0$ K and no stochastic random fields are included, $M_x(H_x)$ is not identical for consecutive field cycles.

In order to elucidate the origin of this nonreproducible response after each field cycle, the normalized magnetization $m_x$ along two consecutive rotating field cycles are shown in Fig. 4 (a). The corresponding micromagnetic states are shown in Fig. 4 (b). The first and second field cycles are denoted with (A) and (B), respectively. According to the $M_x$ ($H_x$) curve of Fig. 4, (a) the states A1 and B1 have an almost identical Mx. Indeed, by comparing the $x$ and $y$ components of the magnetization of state A1 and B1, no difference can be observed in the color code. A detailed view of the magnetization of state A1 is shown in the supplementary material. In order to clearly visualize any difference in the magnetic states, the local difference in the magnetization is calculated according to:

$$Diff = \left|\mathbf{M}_A - \mathbf{M}_B\right| \tag{3}$$

The difference in these two states is shown in the last column of Fig. 4 (b). The scaling color code in the plot showing the difference is arbitrarily chosen, in order to clearly visualize the difference. It can be seen that different residual domains at the top and bottom of the elliptical particle occur. This very small difference in the magnetization is sufficient, for states A2 and B2 to follow completely different branches, as shown in Fig. 4 (a). The corresponding magnetic states are shown in Fig. 4 (b) in the second row. After the $x$ component of the rotating field reaches about -2.5 mT, the two branches merge to two almost identical loops again. However, a detailed inspection of the magnetic states A4 and B4 shows different residual domains again. As a consequence, the third loop (not shown) is again different from the second loop, since the initial state after one full rotation is not exactly the same.

Due to the different $M_x(H_x)$ responses after each cycle, the zero crossings of the $M_x(H_x)$ loop also vary. As a consequence, the very interesting effect can be observed that phase noise occurs, as shown in Fig. 2 (b), in regions that resemble the experimentally obtained phase noise croissant. For larger applied fields, the phase noise vanishes, which is attributed to reproducible and identical initial states after each field cycle, since the field is large enough to annihilate any residual domains or edge domains.

Another very exciting effect can be observed in the micromagnetic simulations, as well as in the experimental data. The hysteresis in the $M_x(H_x)$ loop is larger for the intermediate field state compared to the case of the large field. These results suggest that a minor hysteresis loop can intersect the outer hysteresis, as was also reported in micromagnetic simulations in Ref [21].

To summarize both models, the SW model, and the micromagnetic simulations, a clear origin of the observed phase noise for intermediate fields can be found. It is due to non-reproducible switching of the free layer magnetization. Since the origin of this significant noise source is found, now a disruptive free layer design, that is different to state-of-the-art, is proposed and discussed in the following.

## 4. Sensor with a magnetic vortex state in the free layer

It is well known that thin soft magnetic elements in the micrometer and submicrometer regime form flux closure states, such as Landau patterns [22,23]. These magnetic structures are characterized by parallel magnetization at the boundary of the magnet in order to avoid stray fields due to magnetic surface and volume charges. Phase diagrams of circular $Ni_{80}Fe_{14}Mo_5$ nanomagnets fabricated by high-resolution electron beam lithography show the transition between vortex states and single-domain states [24]. Both analytical theories and micromagnetic simulations were used in order to calculate the phase diagrams from a single domain to vortex states in circular nanodisks [25,26]. Magnetic vortex structures have a topological charge of one, which results in topological protection against a transformation to a magnetic state with topological charge zero, such as the single domain state[2]. Due to the topological protection, vortex structures are stable and robust against thermal fluctuations and external fields. The out-of-plane stray fields of the central vortex core are confirmed by magnetic force microscopy[27,28]. The reversal of the perpendicular magnetic vortex cores with a sinusoidal excitation field as small as 1.5 mT was reported by Waeyenberge et al. [29]. Magnetic vortex oscillations with d.c. spin-polarized currents was reported one year later [30].

The main idea of using flux closure states for sensor applications is to utilize vortex core displacement as a function of the field as the sensor response. If a magnetic field is applied, the magnetic domains that point parallel and anti-parallel to the external field direction increase and decrease in size, respectively. As a consequence, the vortex core is shifted smoothly, and the average magnetization within the free layer is proportional to the external field vector. A schematic hysteresis loop of a vortex state is shown in *Fig. 1* (b). The magnetization component in the field direction increases until the vortex is annihilated at the annihilation field $H_a$. The resulting state is an almost uniform single domain state with the magnetization pointing in the field direction. If the external field is decreased, the vortex core is reestablished at the nucleation field $H_n$. Guslienko et al. developed an analytical model (rigid vortex model) in order to calculate the annihilation field, nucleation field, and the initial susceptibility of magnetic vortex states [31]. The initial susceptibility is an important property for sensor applications, because it defines the sensitivity of the magnetic sensor.

In order to exploit the vortex state in xMR sensors, we utilized pinned layer magnetization with homogeneous magnetization, as shown in *Fig. 1* (b) as reference. Due to the xMR effect, the resistance of the sensor is minimal (maximal) if the free layer magnetization is locally parallel (antiparallel) to the pinned layer magnetization. In Fig. 5 (a), the normalized transfer curve of the experimentally realized vortex sensor (red) is compared with state-of-the-art xMR sensors (blue). A similar structure was used in Ref[[32]] in order to study the influence of edge inhomogeneities on vortex hysteresis curves in magnetic tunnel junctions.

The magnetoresistance ratio (MR) of the elliptical sensor and vortex sensor is $MR_{ellipse}$ = (max(R)-min(R))/min(R) = 10.7% and $MR_{vortex}$ = 5.6%, respectively. The smaller MR ratio of the vortex sensor can be attributed to the thicker free layer, which acts as a parallel resistance and reduces the interfacial GMR effect.

An important property of the vortex sensor is the linear range which is enhanced by a factor of 7 compared to the elliptical sensor  The vortex sensor completes the vortex nucleation at a field of $|\mu_0 H_x| < 23$ mT. Since the sensor consists of 2355 disk structures, thermal effects, as well as structural variations, lead to a distribution of the nucleation fields.

The vortex sensor is also superior to the elliptical sensor concerning hysteresis. While the elliptical sensor has a finite coercive field of $\mu_0 H_c$ = 1.2 mT, the hysteresis of the vortex sensor is vanishingly small and in the range of measurement noise.

The original motivation of the vortex structure was to utilize the stable flux closure configuration as a concept to reduce phase noise. The measurements for all investigated vortex structures show that phase noise indeed vanishes. An example of a phase noise measurement is depicted in Fig. 2 (v1). The vanishing phase noise also agrees with the expected behavior from theory and simulations (see Fig. 2 (v2)), which do not predict any phase noise, either.

Since the dominant noise source, which originates from switching processes in the free layer of elements with shape anisotropy, is not present in vortex sensors, the accuracy is mainly limited by the remaining resistive and magnetic noise, which is described by the total noise power $S_V^{tot}$, common to any magnetic sensor. The total noise power $S_V^{tot}$ is composed of Johnson noise, electric 1/f noise, magnetic 1/f noise, and magnetic white noise [33,34]. In the following, the total noise power of the vortex sensor is compared with the noise of the elliptical sensor outside the phase noise croissant. Details on the noise measurements are described in the method section. Fig. 5 (b) shows $\sqrt{S_V^{tot}}$ as a function of the frequency for the elliptical sensor (blue) and the vortex sensor (red) at zero external field. For both sensors, a clear 1/f noise can be observed until the total noise saturates for high frequencies at the Johnson noise, which is constant with frequency.

Total noise power measurements as a function of the external field correlates with the susceptibility of the sensor as a function of field. Hence, it can be concluded that the dominant noise source has a magnetic origin. The magnetic thermal noise power is expected to scale with [33]

$$S_V^{mag} \propto \frac{1}{NA} \tag{4}$$

where *N* is the number of elements and *A* is the free layer area. In our investigated vortex sensor, the total area is $NA_{vortex} = 2355 \times 3.14 \mu m^2 = 7397 \mu m^2$, which is almost identical to the total area of the elliptical elements, which is $NA_{ellipse} = 1536 \times 4.71 \mu m^2 = 7236 \mu m^2$, and, therefore, directly comparable. It might be interesting to note that the investigated elliptical structures are not arbitrary test structures but sensors that are used in state of the art commercial sensors developed by Infineon AG. Interestingly, the measured noise at 10 Hz of the vortex sensor is smaller by a factor of 7 compared to the elliptical sensor. The significant difference in noise between these two different sensor layouts has to be attributed to different magnetic noise sources due to the different magnetic configurations.

According to the fluctuation-dissipation theorem [33,35], the dissipation is correlated to the imaginary part of the susceptibility $\chi''(f)$ and is proportional to the magnetic noise power as

$$S_V^{mag} \propto \frac{2k_B T \chi''(f)}{\pi N A t_{FL} \mu_0 f} . \tag{5}$$

Here, *N* is the number of sensor elements, *A* the areal of each sensor element, $t_{FL}$ the thickness of the sensor element, *T* the temperature and *f* the frequency. Hence, magnetic processes lead to dissipation, and consequently, a finite $\chi''(f)$ can be related to noise sources. Measurement of $\chi''(f)$ and use of the fluctuation dissipation theorem to extract the noise were performed by Hardner et al. [35]. The obtained noise values agreed well with the directly measured total noise power. An indication of the low noise of the vortex sensor can be found in the small hysteresis of this sensor design, which in turn correlates to a small $\chi''(f)$. In contrast to the vortex structure, any misalignment of the long axis of the ellipses perpendicular to the field direction leads to hysteresis.In the elliptical sensor, all spins—also the surface spins—rotate by applying external fields, which might lead to local pinning and residual domains, contrary to the vortex states

The total magnetic noise power scales with the magnetoresistance as $S_V^{mag} \propto MR^2$. Hence, if, for equivalent sensors, the MR ratio is increased, the noise increases, as well. However, the larger noise does not imply that the minimum fields that can be detected are larger for sensors with higher MR values. The reason is that with increasing MR, in addition to the noise, the sensitivity $\frac{dV}{\mu_0 dH_{ext}}$ also increases. The sensitivity relates the change of sensor output to the change of the applied magnetic field $\mu_0 H_{ext}$. Hence, a more meaningful property to compare different sensors is the detectivity *D*

$$D = \sqrt{S_{\mu_0 H}} = \left(\frac{\mu_0 dH_{ext}}{dV}\right)\sqrt{S_V} \quad \left[T / \sqrt{Hz}\right] \tag{6}$$

which relates the sensor noise to the sensor sensitivity. The detectivity is proportional to the signal-to-noise ratio, as shown in the method section. Hence, a smaller detectivity allows to detect smaller magnetic fields. In Fig. 5 (c), the detectivity of the two investigated sensors is compared. The detectivity at 10 Hz is D ~ 8 $nT / \sqrt{Hz}$ for the elliptical and D ~ 20 $nT / \sqrt{Hz}$ for the vortex sensor, respectively. Hence, even if the linear range of the vortex sensor is larger by a factor of 7 compared to the elliptical sensor, the detectivity is only smaller by a factor of 2.5. For comparison, we introduce the figure of merit (FM), which relates the linear range to the detectivity, as

$$FM(f) = \frac{linear\,range}{D(f)} \tag{7}$$

The FM is larger for the vortex sensor by a factor of 2.7.

## 5. Summary and discussion

Within this work, it is shown that the noise of state-of-the-art xMR sensors with an elliptical free layer design strongly depends on the rotating field amplitude. For certain ratios of parallel and perpendicular components of the rotating field, the phase noise increases by more than one order of magnitude. For small applied fields, the free layer magnetization reversibly tilts out from the easy axis, and for larger applied fields, the free layer magnetization reproducibly irreversibly switches, leading to small noise levels. However, for field amplitudes that are close to the SW astroid, the free layer magnetization switches in a non-reproducible way. Two mechanisms can be attributed to the non-reproducible switching. The first process, which even predicts noise for perfect single domain element, occurs due to finite temperature. Here, the free layer magnetization has a certain probability for thermally activated switching to the opposite equilibrium state. For angles between the free layer and the pinned layer that are not exactly 90°, the sensor response is different for the up and down states. As a consequence, phase noise occurs. The second noise source for fields close to the SW astroid is obtained even at zero temperature.This noise source does not require any stochastic input. With micromagnetic simulations, consecutive hysteresis loops show for fields close to the SW astroid that the final state is slightly different after each field cycle. As a consequence, the sensor responses are different in each field cycle. Field conditions that lead to heavy noise also lead to the largest hysteresis of the sensor. Interestingly, the hysteresis is not the largest for the largest applied field, as shown by experimental data, as well as by micromagnetic simulations[21].

A disruptive sensor layout is presented that avoids the aforementioned problems by rotational symmetry of the free layer magnetization that forms a flux-closed vortex configuration. Micromagnetic simulations, as well as experiments, show a basically hysteresis-free sensor response in the vicinity of the zero line crossing of the transfer curve. The reproducible vortex core motion as a function of applied fields overcomes the phase noise problem. Besides the phase noise, the total noise power $S_V^{tot}$ is significantly reduced for vortex sensors compared to elliptical sensors. The linear range of the vortex sensor can be tuned by the ratio of the free layer thickness to the diameter, and it is almost one order of magnitude larger than for elliptical sensors. A large linear range of sensors is of utmost importance for wheel speed applications, because a higher dynamic range enables less effort for the adjustment of the operational working point. Due to the reduced noise of vortex sensors, the detectivity, which describes the minimum detectable field, is only smaller by about a factor of two compared to state-of-the-art sensors and leads to a larger figure-of-merit for the vortex sensor by a factor of 2.7.

This work clearly demonstrates that the presented flux closure vortex state outperforms state-of-the-art GMR and TMR sensors, which are widely used in applications worldwide. This work demonstrates the potential advantages of topologically protected structures with unique magnetization states and might be guiding the way for commercial applications of topologically protected magnetic states [36].

Support from CD-laboratory AMSEN (financed by the Austrian Federal Ministry of Economy, Family and Youth, the National Foundation for Research, Technology and Development), the FWF – SFB project F4112-N13, the Vienna Science and Technology Fund (WWTF) under grant MA14-044, and the Advanced Storage Technology Consortium (ASTC) is acknowledged. The computational results presented were achieved using Vienna Scientific Cluster (VSC).

## 6. Methods

### Calculation of phase noise

The phase noise $\sigma_\tau$ of the magnetic sensor under the action of an elliptical rotating field is defined as:

$$\sigma_\tau = \frac{\sqrt{\frac{1}{N-1}\sum_{i=1}^{N}\left(\tau_i^{period} - \overline{\tau}\right)^2}}{\overline{\tau}} \tag{8}$$

where $\overline{\tau}$ is the average period length:

$$\overline{\tau} = \frac{1}{N}\sum_{i=1}^{N}\tau_i^{periode} \tag{9}$$

and $\tau_i^{periode}$ is the $i$ -th period:

$$\tau_i^{periode} = t_n - t_{n-1} \tag{10}$$

where $t_n$ is the time of the $n$th zero crossing of either the rising or falling edge of the resistance $R$.

### Implementation of SW model at finite temperature

In order to implement thermal effects, a model based on the transition theory is implemented [37]. The magnetization dynamic is not described by solving the Landau-Lifshitz-Gilbert equation, which leads to a decrease of energy. Due to finite temperature, energy barriers for switching may be overcome by thermal activation. The energy barrier is given by the energy difference between the lower energy maximum and energy of the current minimum. The relaxation time for the switching from one stable magnetization state to the other is given by the Arrhenius-Néel law:

$$\tau = \tau_0 e^{-\frac{\Delta E}{k_B T}} \tag{11}$$

where $\tau_0$ is a characteristic time constant, the attempt period (its reciprocal is the attempt frequency $f_0$), which is on the order of $10^9$ to $10^{11}$ Hz [38] , $k_B$ is the Boltzmann constant, $T$ is the temperature, and $\Delta E$ is the energy barrier separating the two states. The probability for switching $p$ from one state to the other within the time $\tau_m$ is given by

$$p = 1 - e^{-\frac{\tau_m}{\tau}} \tag{12}$$

For most field amplitudes, once the particle switches from one state to the other, it will remain in that other state, because the energy barrier for the inverse process is much higher. However, if the field vector is near the hard axis cusps of the SW astroid but on the inside, multiple switching events are possible, because the energy levels for both states are very similar. Once the particle switches, as long as the field remains near the cusp, there is a possibility it will simply switch back. We can define the time constants and probabilities for both switching processes as

$$\tau_{\uparrow\downarrow} = \tau_0 e^{-\frac{\Delta E_{\uparrow\downarrow}}{k_B T}} \tag{13}$$

$$\tau_{\downarrow\uparrow} = \tau_0 e^{-\frac{\Delta E_{\downarrow\uparrow}}{k_B T}} \tag{14}$$

where $\Delta E_{\downarrow\uparrow}$ and $\Delta E_{\uparrow\downarrow}$ is the energy barrier height within the down state and up state, respectively. Consequently, the respective transition probabilities are given by

$$p_{\uparrow\downarrow} = 1 - e^{-\frac{\tau_m}{\tau_{\uparrow\downarrow}}} \tag{15}$$

$$p_{\downarrow\uparrow} = 1 - e^{-\frac{\tau_m}{\tau_{\downarrow\uparrow}}} \tag{16}$$

In order to simulate the thermally activated process, the time $t$ in Eq. (1) is discretized in equal time steps $\tau_m$. Within each time step, the switching probability $p$ according to Eq. (12) is calculated. The probability is compared with a random number $r$ between $0 \le r \le 1$. If $p > r$, the particle is reversed and set into the new equilibrium state.

**Experimental setup to measure domain phase noise**

Phase noise measurements were performed in a 3-axis Helmholtz coil from Micromagnetics Inc. with a homogeneous magnetic region of (30x30x30) mm³. The size of the specimen with the bonded GMR device is 2.79 mm x 1.79 mm. The GMR device is stimulated with magnetic fields of sinusoidal shape in the *x* and *x* direction, and the voltage change was read via 4-point measurement ($I_{force}$ = 180 µA). For proper phase noise determination, a high-speed data acquisition card of NI with 8 simultaneous inputs and 2 analog outputs was used. The sample rate was variable with 5000 data points per period. Hence, the stimulated signal at 50 Hz used a sample time of 4 µs. To ensure acceptable statistics, 300 periods were measured, and the background noise was below 0.01 %.

### Experimental setup of total noise power measurement

The device under test (DUT) is supplied by a battery to reduce external noise sources. To regulate the applied voltage on the DUT, a potentiometer is connected in series, and the current is monitored with a multimeter. The fully differential preamplifier is based on the work of Scandurra et al.[39]. It uses two symmetrical JFETs for a very low equivalent input voltage noise, followed by two operational amplifiers in the first stage, resulting in an amplification of 100. A second-stage operational amplifier removes the big DC component produced by the JFETs and amplifies the input signal by a factor of 100. The total amplification of both stages gives a factor of $10^4$. Good performance can be reached in the range of 4 Hz to 10 kHz. A DSP Lock-in Amplifier (LIA) SR830 from Stanford Research Systems is used to obtain the amplified noise from the preamplifier. It produces a noise value with a bandwidth of 0.01 Hz.

To perform field measurements, the DUT is placed in the center of a GMW 3470 dipole electromagnet, generating a homogeneous magnetic field at the sensor position. The current for the electromagnet is regulated by a bipolar operational amplifier (BOP from the KEPCO Company). Supply and monitoring of the sensor are done by a Keithley 2400 sourcemeter.

### Fabrication of the sensor stack

The sensor stack is fabricated by physical vapor deposition (PVD). The magnetic stack is structured by photolithography and ion beam etching. The sensor stack consists of a synthetic antiferromagnetic (SA) reference system (pinned layer) of two CoFe Layers that are antiparallelly coupled via an Ru layer. The SA is pinned to an antiferromagnetic PtMn layer. A Cu layer is used as a spacer. The free layer consist of CoFe-based alloys. The sensor stack is contacted by vias, as shown schematically in Fig. 2 (v3), that are connected through the substrate to the reference system. The current flow is in plane of the sensor stack (CIP).

### Contributions

D.S., A. S, J. Z, W. R., H.B. conceived the vortex concept for low noise sensing. C.P. prepared the samples. A.S., H.W., C. P., C. H., H. B. designed and performed the experiments and analysed the experimental data. A.B-H. performed and evaluated the micromagnetic simulations. D.S., A. B-H., C. V., F. B., C. A. and T.S. developed and improved the micromagnetic code. D.S., A.B-H., A. S., C.P., J.Z., S.L., W.R. and H.B. interpreted the results. D.S., A. B-H. and H. B. wrote the manuscript with input from all coauthors. All authors discussed the results and commented on the manuscript.

**Acknowledgement**

The financial support by the Austrian Federal Ministry of Science, Research and Economy and the National Foundation for Research, Technology and Development as well as the Austrian Science Fund (FWF) under grant F4112 SFB ViCoM is gratefully acknowledged. The computational results presented have been achieved using the Vienna Scientific Cluster (VSC).

**Competing financial interests**

The authors declare no competing financial interests.

**Corresponding author**

Correspondence to: Dieter Suess

Fig. 1: Schematic layout and transfer characteristics of spin valve sensor structures consisting of a magnetically pinned layer (violet), a spacer layer (orange), and a magnetically free layer (blue). (a) State-of-the-art xMR sensor; (b) Vortex state structure .

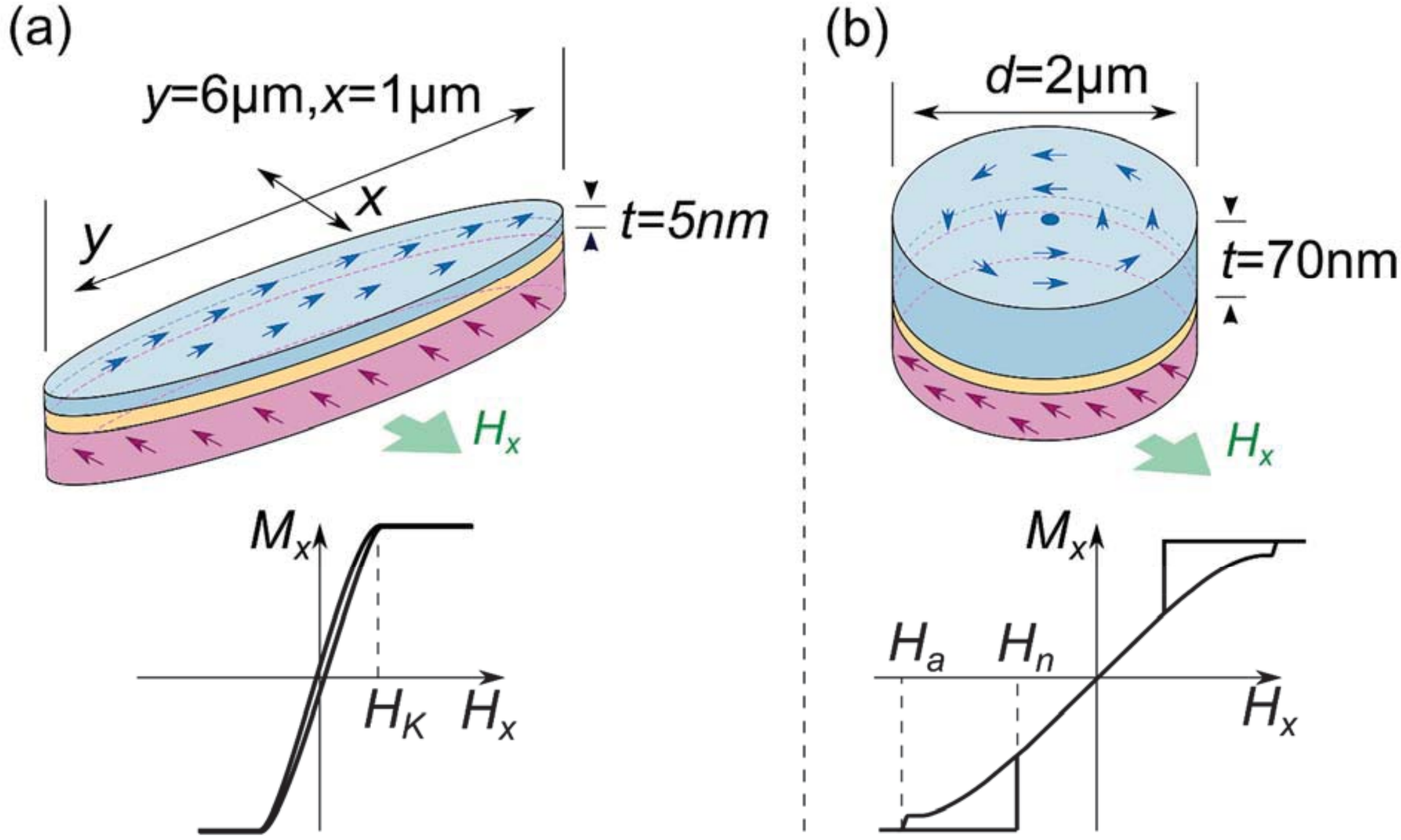

Fig. 2: (e1-e3) Falling edge phase noise of elliptical GMR sensor with free layer dimensions of $y$ = 6 µm, $x$= 1 µm and thickness $t$ = 5 nm (geometry of state of the art commercially used sensor). (e1) Measurement of 16 electrically connected, elliptical CoFe transducer elements. (e2) Micromagnetic simulations of the phase noise of a single element (e3). Phase noise obtained from the Stoner-Wohlfarth model at T = 300 K.

(v1-v2) Phase noise of a GMR vortex sensor of diameter $d$ = 2 µm and thickness $t$ = 80 nm. (v1) Experimentally obtained phase noise of electrically connected 1752 disk-shaped elements (v3) The vortex sensor is shown with the contacts (vias) schematically.

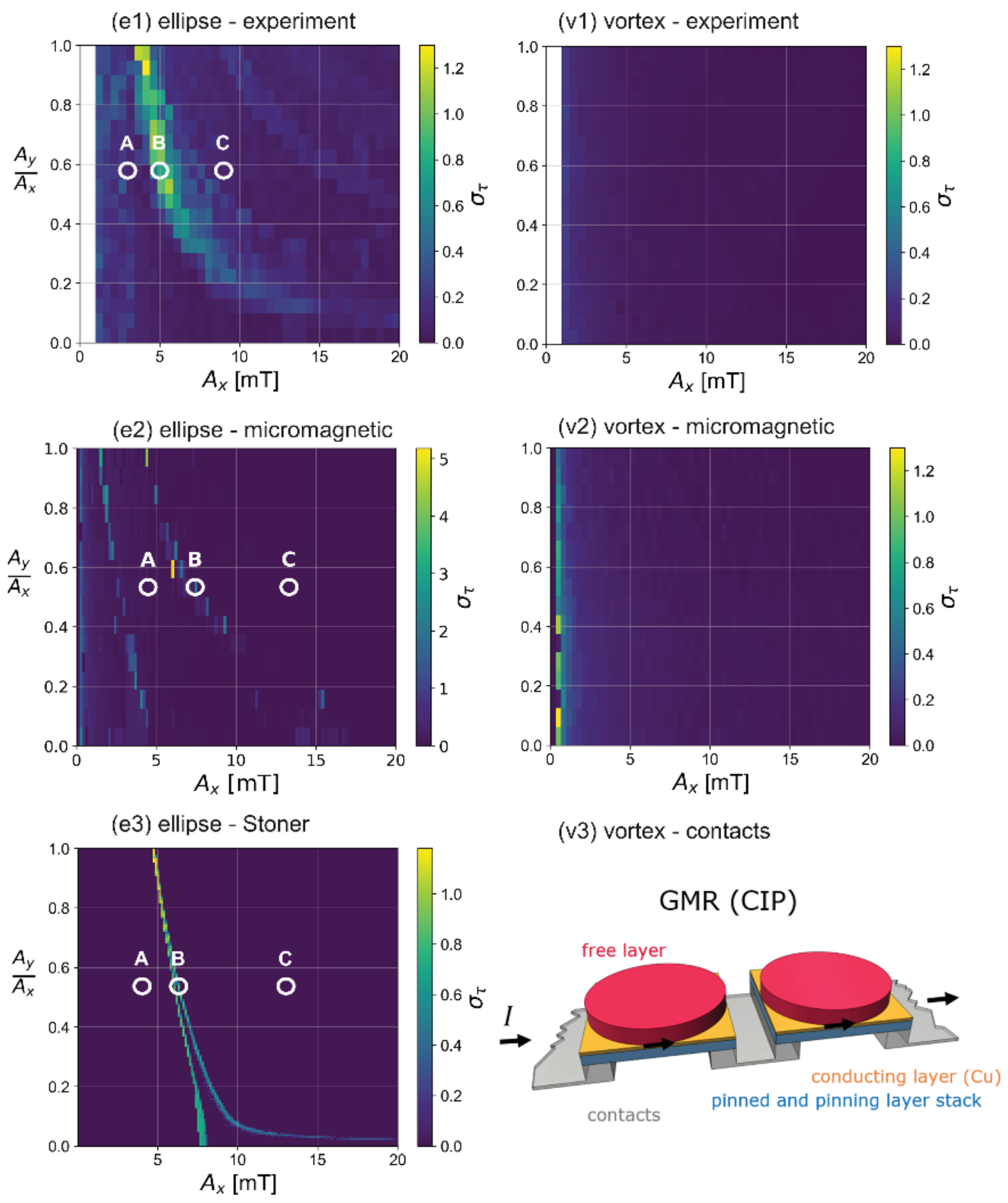

Fig. 3: Comparison of $M_x(H_x)$ for the measurement (green), the Stoner–Wohlfarth model (blue), and the micromagnetic model (red). Three different characteristic rotating field strengths are shown. (a) Fields smaller than the phase noise croissant: Stoner–Wohlfarth: $A_x$ = 4.035 mT, $A_y$ = 2.166 mT; micromagnetic: $A_x$ = 4.5 mT, $A_y$ = 2.4 mT; measurement: $A_x$ = 3 mT, $A_y$ = 1.737 mT. (b) At the phase noise croissant: Stoner–Wohlfarth: $A_x$ = 6.316 mT, $A_y$ = 3.391 mT; micromagnetic: $A_x$ = 7.43 mT, $A_y$ = 3.967 mT; measurement: $A_x$ = 5 mT, $A_y$ = 2.895 mT. (c) Fields larger than the phase noise croissant, $A_x$ = 13.02 mT, $A_y$ = 6.993 mT; micromagnetic: $A_x$ = 13.32 mT, $A_y$ = 7.102 mT; measurement: $A_x$ = 9 mT, $A_y$ = 5.211 mT.

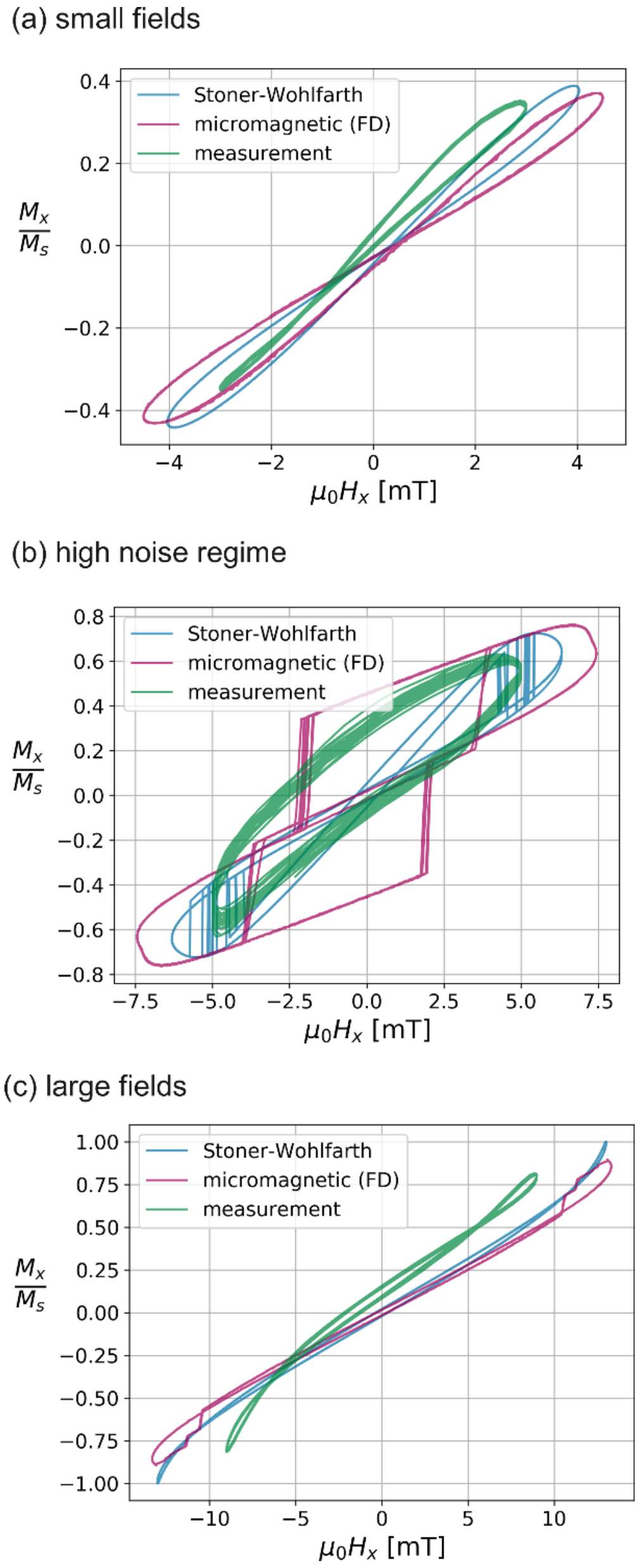

Fig. 4: (a) $M_x$ component of the elliptical element of Fig. 2 as a function of $H_x$ of a rotating field. The first hysteresis cycle is marked by the states A1-A4 (red). The second hysteresis cycle is marked by the states B1-B4 (green). (b) The magnetic *x* and *y* component of the magnetization of the states A1 to A4 and B1 to B4 are color-coded. The last column (Diff) shows the magnitude of the difference vector of the corresponding state A and state B. The color code in the Diff- column is blown up to clearly visualize regions with large difference.

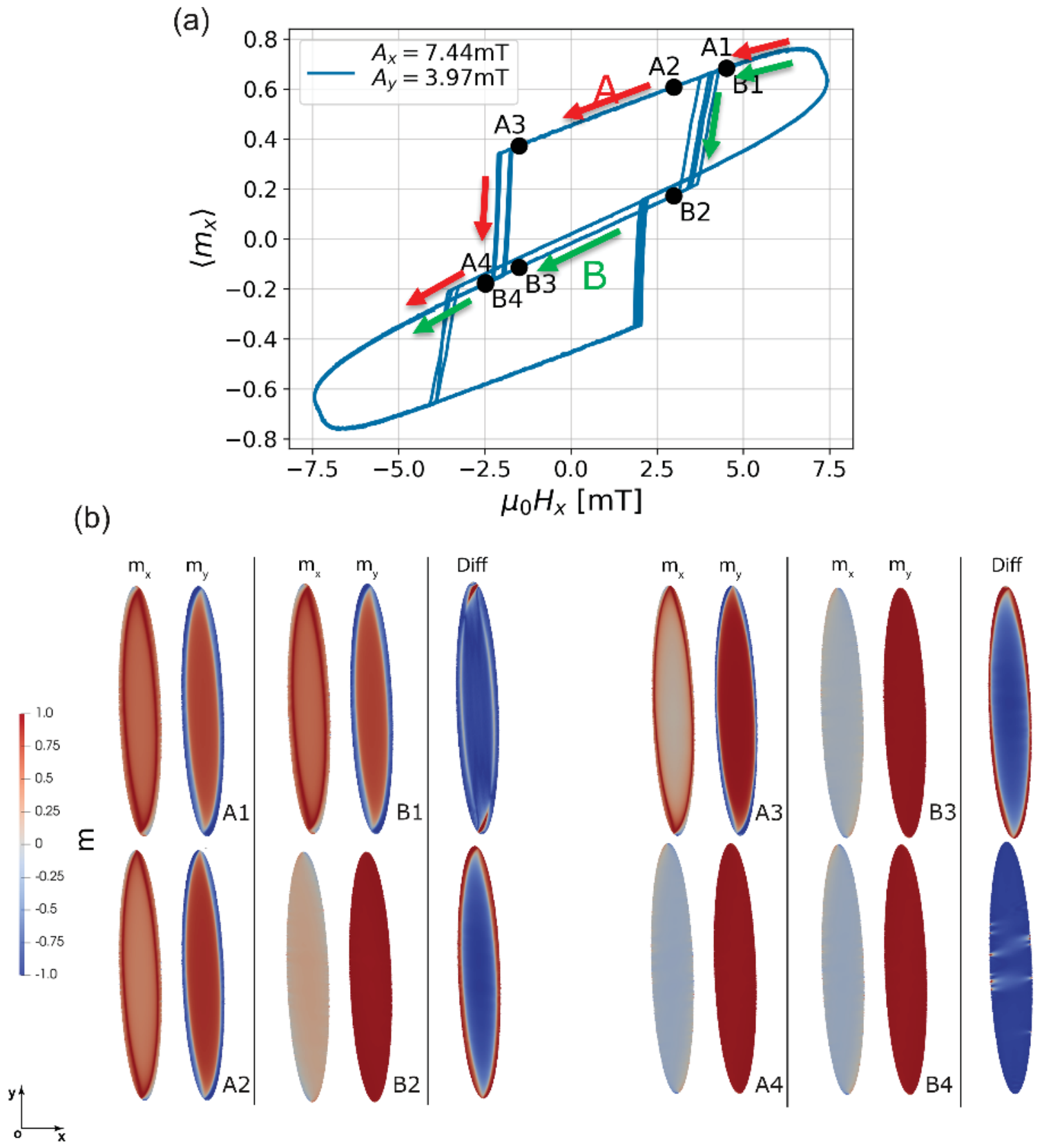

Fig. 5: Comparison of sensor performance of an elliptical GMR sensor (state of the art commercially used sensor) ($x$ = 1 µm , $y$ = 6 µm, $t$ = 5 nm, array of 1536 elements) and a disk-shaped vortex GMR sensor (d = 2 µm, $t$ = 70 nm, array of 2355 elements) with a CoFe based free layer. The external field is applied along the reference magnetization. (a) The normalized transfer curve, showing a linear range of the vortex sensor, which is larger by a factor of about 7 than for the elliptical sensor. (b) Comparison of the frequency-dependent noise for a supply voltage of 2.5 V. (c) Comparison of the detectivity (field noise).

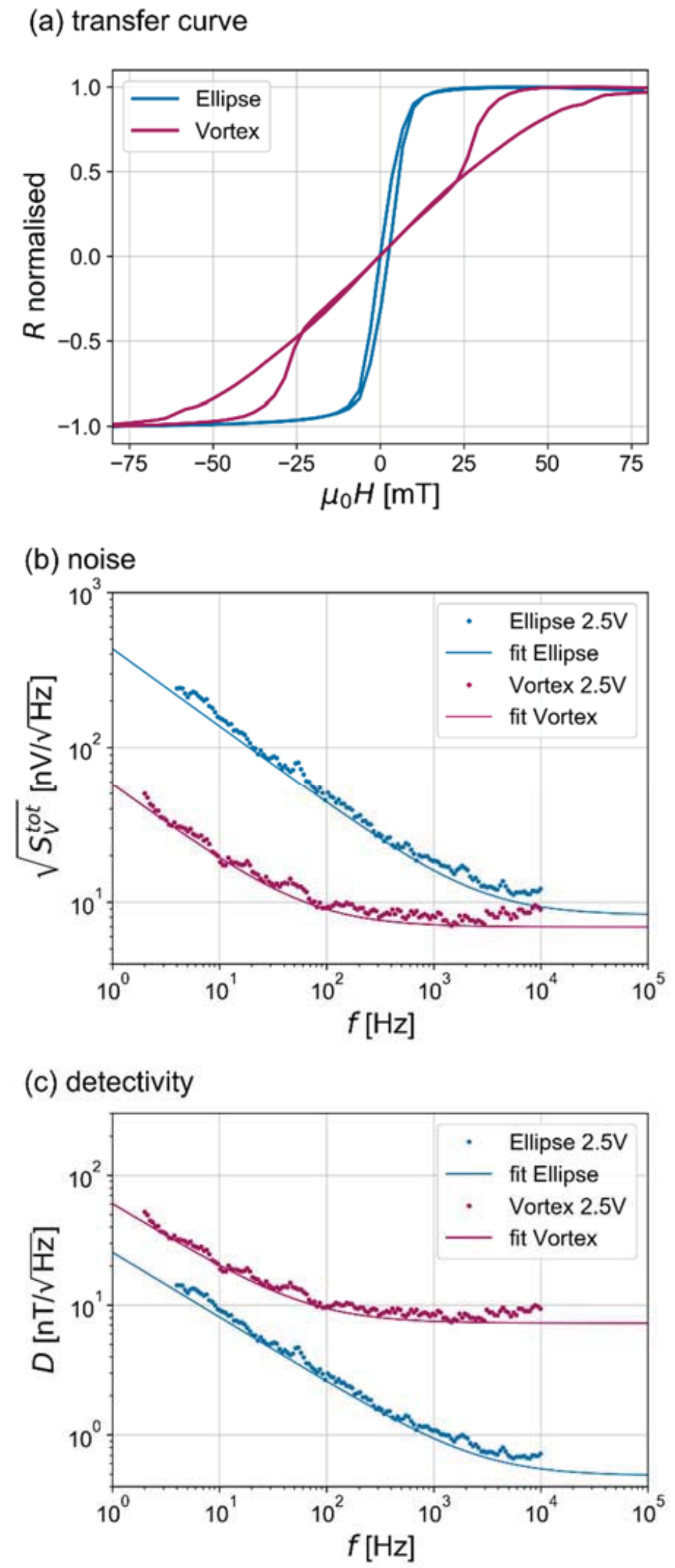

# Supplementary Information - Topologically Protected Vortex Structures to Realize Low Noise Magnetic Sensors

**Dieter Suess[1,*], Anton Bachleitner-Hofmann[1], Armin Satz[2], Herbert Weitensfelder[1], Christoph Vogler[1], Florian Bruckner[1], Claas Abert[1], Klemens Prügl[4], Jürgen Zimmer[3], Christian Huber[2], Sebastian Luber[3], Wolfgang Raberg[3], Thomas Schrefl[5], and Hubert Brückl[5]**

[1]University of Vienna, Christian Doppler Laboratory, Faculty of Physics, Physics of Functional Materials, Währinger Straße 17, 1090 Vienna, Austria
[2]Infineon Technologies AG, Siemensstraße 2, 9500 Villach, Austria
[3]Infineon Technologies AG, Am Campeon 1-12, 85579 Neubiberg, Germany
[4]Infineon Technologies AG, Wernerwerkstrasse 2, 93049 Regensburg, Germany
[5]Center for Integrated Sensor Systems, Danube University Krems, Viktor Kaplan Str. 2 E, 2700 Wiener Neustadt, Austria
[*]corresponding.author: dieter.suess@univie.ac.at

## ABSTRACT

In this supplementary material details about the experimental grain structure, the simulation of vortex structures with edge roughness, the origin of phase noise in Stoner Wohlfarth particles and critical fields as well as the connection between detectivity and minimum measurable magnetic fields are given.

## Experimental

### Granular structure of the free layer

A typical free layer of state of the art sensors is shown in fig. (1), which is obtained by a transmission electron microscopy using a FEI Tecnai F20 at TU-Wien/USTEM. In order to extract the grain size of the CoFe free layer a planar view is shown. The average grain size is about 5nm which is sufficiently small leading to a nano-crystalline material with soft magnetic properties[1].

## Simulation

### Edge roughness

To create rectangular and elliptical finite difference models with parameterizable edge roughness, the edges of the base shape are discretized in segments of length $\Delta s$. While this is trivial for rectangular shapes, the segments for the elliptical shapes are found iteratively, where the segment joints are given by

$$\boldsymbol{r}(\varphi_\mathrm{i}) = \begin{pmatrix} b\cos\varphi_\mathrm{i} \\ a\sin\varphi_\mathrm{i} \end{pmatrix} \tag{1}$$

where $a$ is the semi major axis, $b$ is the semi minor axis, and $\varphi_\mathrm{i}$ are the phase parameters corresponding to the respective segment joints. To iteratively find the next segment joint and the corresponding phase parameter, $\varphi_{\mathrm{i}+1}$ is increased until

$$|\|\boldsymbol{r}(\varphi_\mathrm{i}) - \boldsymbol{r}(\varphi_{\mathrm{i}+1})\| - \Delta s| < c\cdot\Delta s \qquad \varphi_\mathrm{i}, \varphi_{\mathrm{i}+1} \in [0, 2\pi], \qquad \varphi_{\mathrm{i}+1} > \varphi_\mathrm{i} \tag{2}$$

is satisfied, where $c$ was taken as $10^{-4}$. This is an approximation, because segments are measured on direct lines between joints, rather than on the actual arc length of the ellipse. Because measuring the distance over the arc length would require solving elliptic integrals iteratively, which cannot be done analytically anyway, and because the discretization length is much smaller than the semi axes $\Delta s << \min(a,b)$, which means that over the distance of $\Delta s$, the edge is nearly flat.
For both rectangle and ellipse, each of the segment joints is then displaced along the edge normal $\boldsymbol{n}(\varphi_\mathrm{i})$ by a distance $a_\mathrm{i}$, where $a_\mathrm{i}$ are random values, drawn from a normal distribution with expectation value of 0 and a standard deviation of $\sigma$. The distorted position of the segment joints is then given by

$$\boldsymbol{r}'(\varphi_\mathrm{i}) = \boldsymbol{r}(\varphi_\mathrm{i}) + a_\mathrm{i}\boldsymbol{n}(\varphi_\mathrm{i}). \tag{3}$$

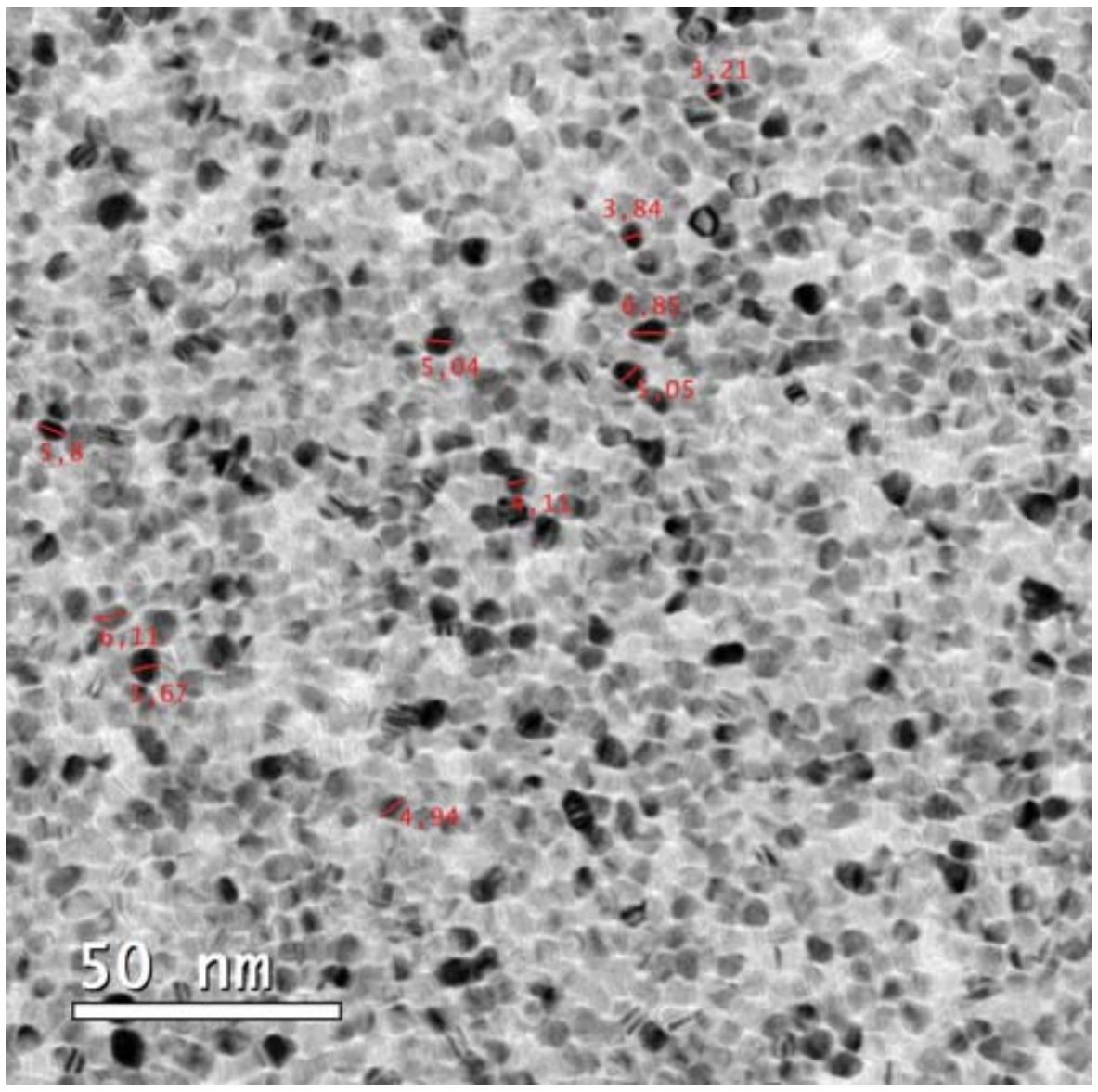

**Figure 1.** TEM plane-view micrograph of a typical sensor stack with $t_{\mathrm{Cu}}$ = 2.25 nm, $t_{\mathrm{CoFe}}$ = 15 nm

For the rectangular shapes (fig. 2b), the edge normal is trivially given by the unit vectors $+\boldsymbol{e}_x$, $+\boldsymbol{e}_y$, $-\boldsymbol{e}_x$ and $-\boldsymbol{e}_y$ for right, top, left and bottom edge respectively. For the elliptical shapes (fig. 2a), the edge normal is given by

$$\boldsymbol{n}(\varphi_\mathrm{i}) = \frac{1}{\|\boldsymbol{n}'(\varphi_\mathrm{i})\|}\boldsymbol{n}'(\varphi_\mathrm{i}) \tag{4}$$

$$\boldsymbol{n}'(\varphi_\mathrm{i}) = \begin{pmatrix} a\cos\varphi_\mathrm{i} \\ b\sin\varphi_\mathrm{i} \end{pmatrix} \tag{5}$$

## Comparison annihilation and nucleation field between simulation and theory

If the external field is increased, the vortex core moves towards the disk edge, but is topologically stabilized as long as it remains inside the disk. The critical field value at which the vortex is pushed out of the disk, is called the annihilation field $H_a$. In the $M(H)$ curve, the annihilation can be identified by a distinct jump towards saturation (see figs. 3 and 4, (a) to (b)), because the magnetization changes to a uniform pattern where the magnetic moments are aligned with the applied field. If the applied external field is decreased again, at some point the vortex state will start re-forming. The nucleation process is typically not as sudden as the annihilation (see figs. 3 and 4, (d) to (f)). The field value at which a vortex state is reached again (fig. 4f) is called the nucleation field $H_n$. Analytically, the critical fields of a vortex disk can be estimated by the rigid vortex model[2,3]. The model is attributed 'rigid' because energy terms are calculated under the assumption that the fieldless, remanent state, where the vortex core resides in the center of the disk, is rigidly displaced under the influence of an external field, and the magnetization is always aligned in concentric circles around the vortex core, even for fields near the annihilation field $H_\mathrm{a}$. The model predicts critical fields of

$$h_\mathrm{a}(\beta,R) = 2a(\beta,R) \tag{6}$$

$$h_\mathrm{n}(\beta,R) = \frac{F_1(\beta)}{2} - 2\left(\frac{L_\mathrm{ex}}{R}\right)^2 \tag{7}$$

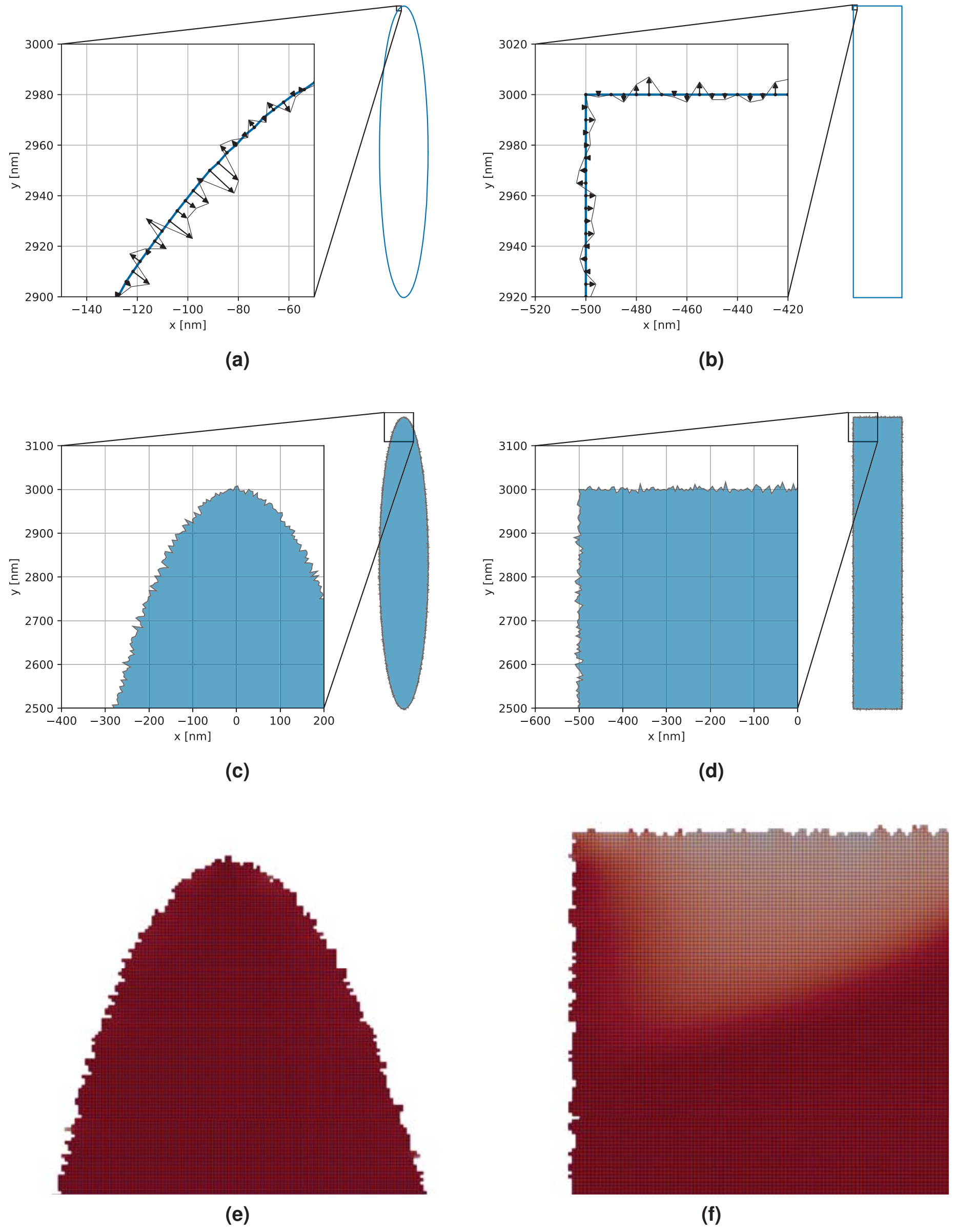

**Figure 2.** (a) & (b) To obtain micromagnetic models with edge roughness, elliptic and rectangular edges are discretized into segments of equal length $\Delta s$. The respective joints are then displaced along the edge normal by random values $\Delta r$ drawn from a normal distribution with expectation value of zero, and a standard deviation of $\sigma$. The parameters of the edge roughness are then given by $\Delta s \times \sigma$. (c) & (d) outlines of the particles with distorted edge for a roughness of $5\,\mathrm{nm} \times 5\,\mathrm{nm}$. (e) & (f) resulting finite difference model for above shapes and a mesh of cubic cells with dimensions $5\,\mathrm{nm} \times 5\,\mathrm{nm} \times 5\,\mathrm{nm}$. Cells are included in the model if the cell center is within the calculated outlines (see (c) & (d)) and excluded otherwise.

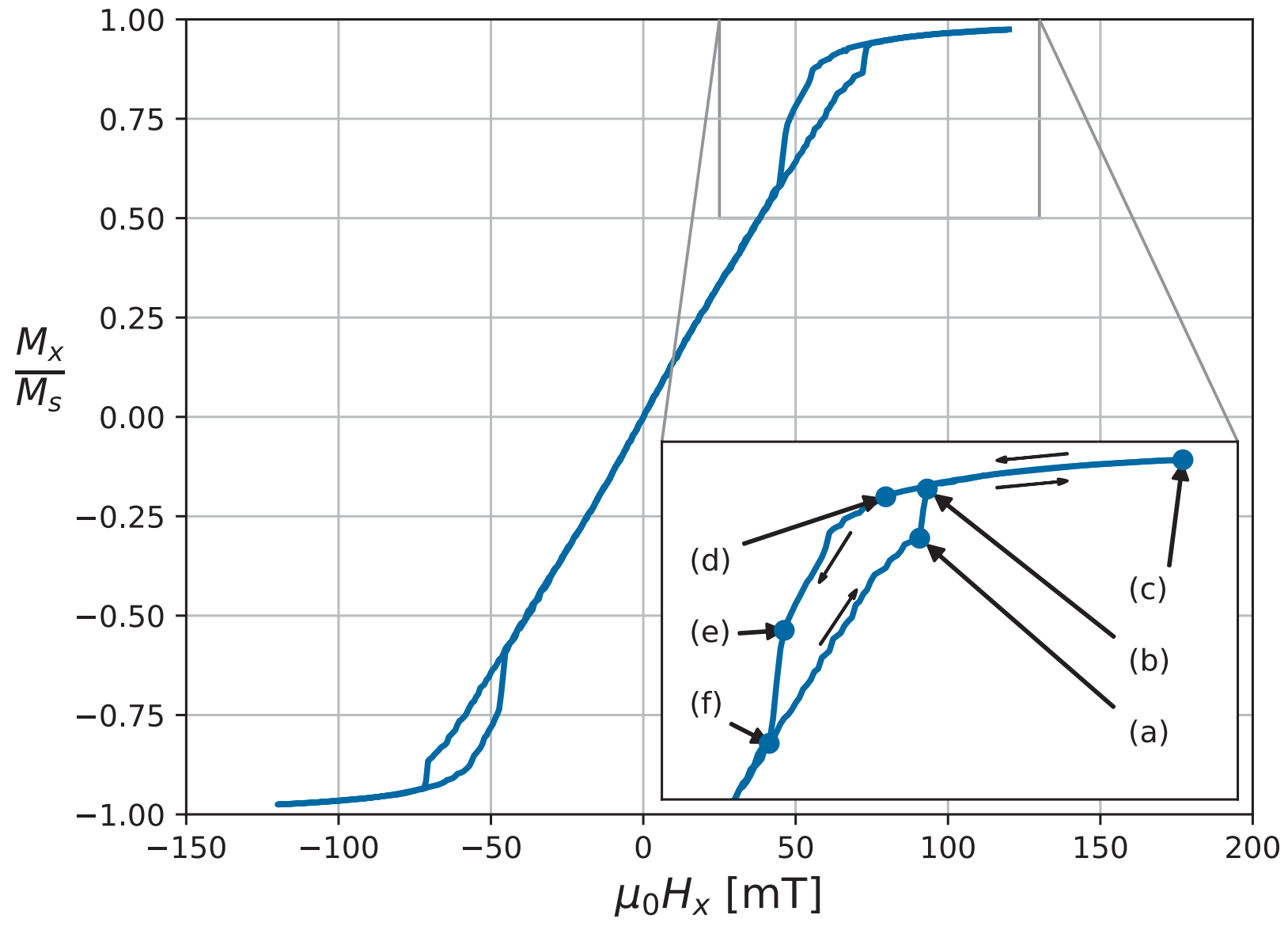

**Figure 3.** Simulated $M_\mathrm{x}(H_\mathrm{x})$ curve of a soft-magnetic CoFe disk with $D = 1600\,\mathrm{nm}$ and $t = 65\,\mathrm{nm}$.

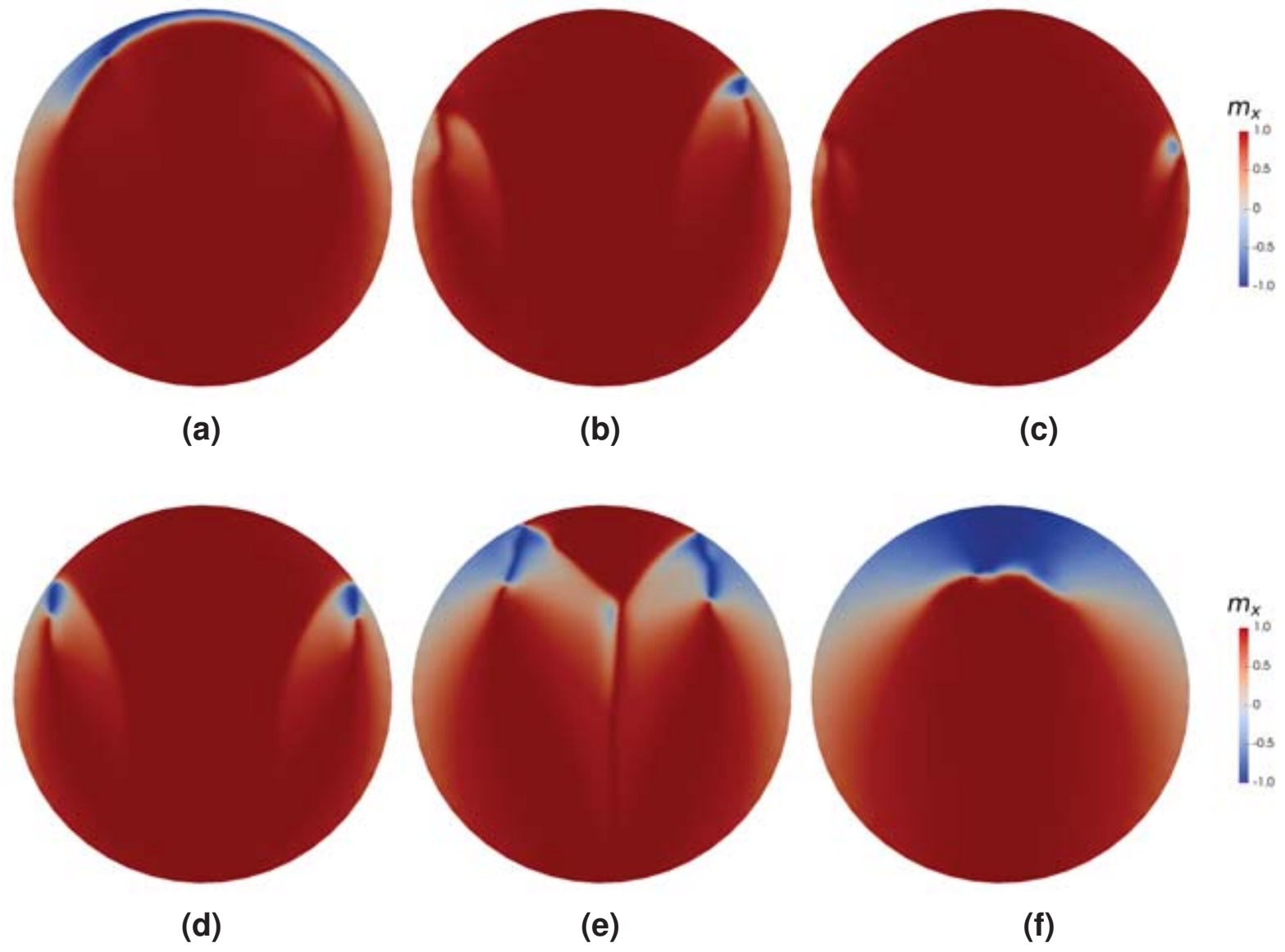

**Figure 4.** Vortex magnetization patterns near the critical fields $H_\mathrm{an}$ (a-c) and $H_\mathrm{n}$ (d-f). The corresponding points on the $M_\mathrm{x}(H_\mathrm{x})$ curve are marked in fig. 3

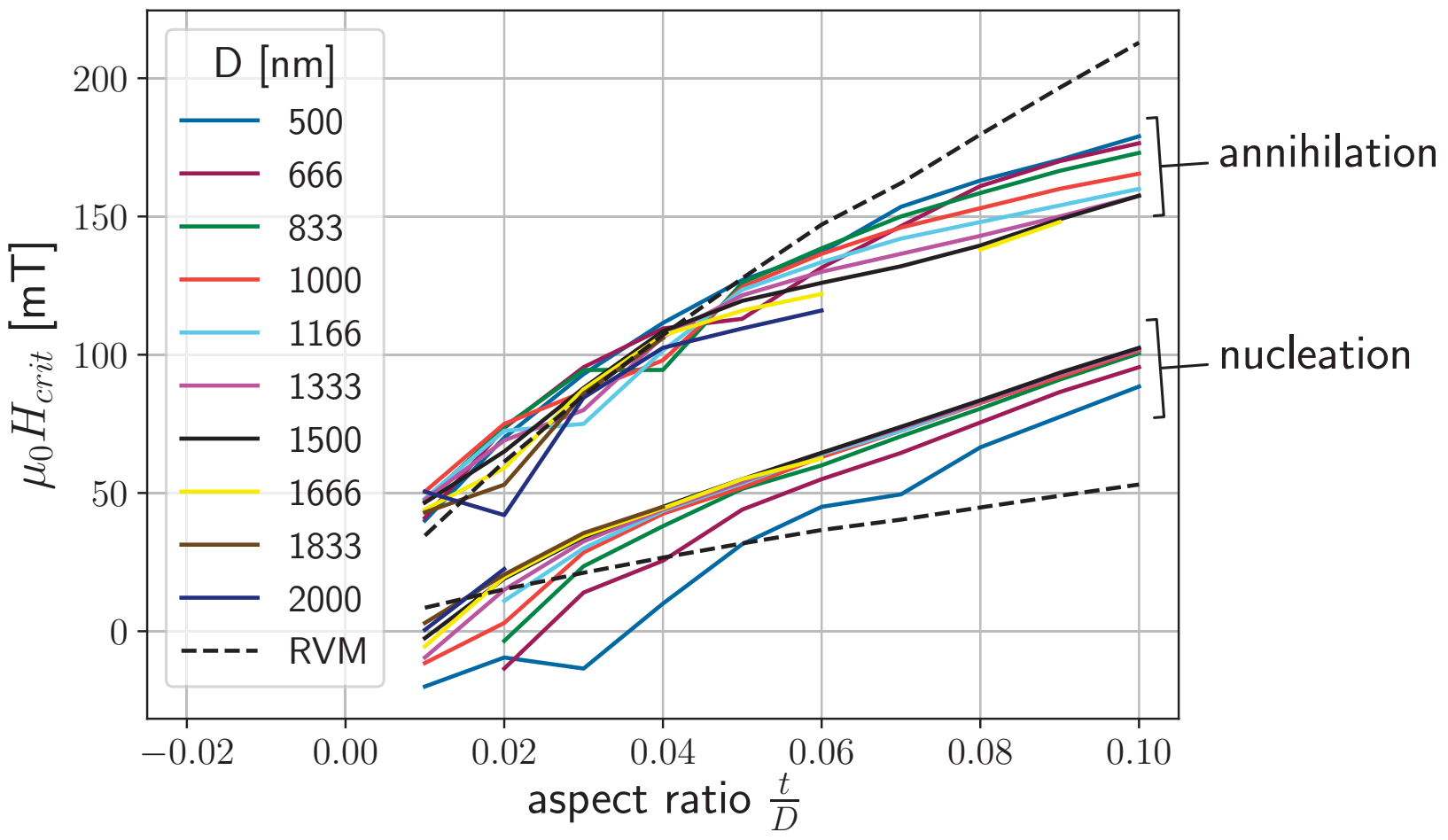

**Figure 5.** Nucleation- and annihilation fields over aspect ratio of thickness to diameter $\frac{t}{D}$. Disks with diameters from 500 nm to 2 µm have been simulated. The dashed lines are the critical fields predicted by the rigid vortex model[2].

where $H_a = \mu_0 M_s h_a$, $H_n = \mu_0 M_s h_n$, $\beta$ is the aspect ratio given by the quotient of disk thickness $t$ and disk radius $R$ and $L_{ex}$ is the exchange length given by

$$L_{ex} = \sqrt{\frac{2A}{\mu_0 M_s^2}} \tag{8}$$

where $A$ is the exchange stiffness. The remaining terms in (6) are given by

$$a(\beta, R) = F_1(\beta) - \frac{1}{2}\left(\frac{L_{ex}}{R}\right)^2 \tag{9}$$

$$F_1(\beta) = \int_0^\pi \frac{dt}{t} f(\beta t) J_1^2(t) \tag{10}$$

$$f(x) = 1 - \frac{1-e^{-x}}{x}. \tag{11}$$

$J_1$ is the Bessel function of the first kind.

Figure 5 shows finite difference simulations of vortex disks with diameters from 500 nm to 2 µm and respective aspect ratios $\frac{t}{D}$ of 0.01 to 0.1. The critical fields predicted by the rigid vortex model are drawn as dashed lines. For very thin disks, the results of the micromagnetic simulations and the rigid vortex model are in very good agreement. As the samples get thicker, the assumption that the vortex core can be described as a localized pillar is no longer accurate, which is reflected in the deviation from rigid vortex model to simulation.

## Zero crossing for tilted SW astroid

In this section we will derive an analytic equation for the field $H_x$ that leads to $M_x = 0$ for a system where the direction of the easy axis of the SW particle and the magnetization direction in the reference system has a finite angle. In the following magnetization and field are expressed in cartesian coordinates, and all fields are of the form

$$\boldsymbol{H}_{ext} = \begin{pmatrix} A_x \cdot \cos t \\ A_y \cdot \sin t \end{pmatrix} \tag{12}$$

Easy axis tilt angles are assumed to be small ($\leq 10°$) and are measured from the positive y-axis in a manner such that a positive tilting angle $\zeta$ leads to an easy axis that lies in the second and fourth quadrant of the cartesian coordinate system. As discussed in the paper the $M_x(H_x)$ loop for a Stoner-Wohlfarth particle with a nonzero tilting angle $\zeta$ are different for up and down state.

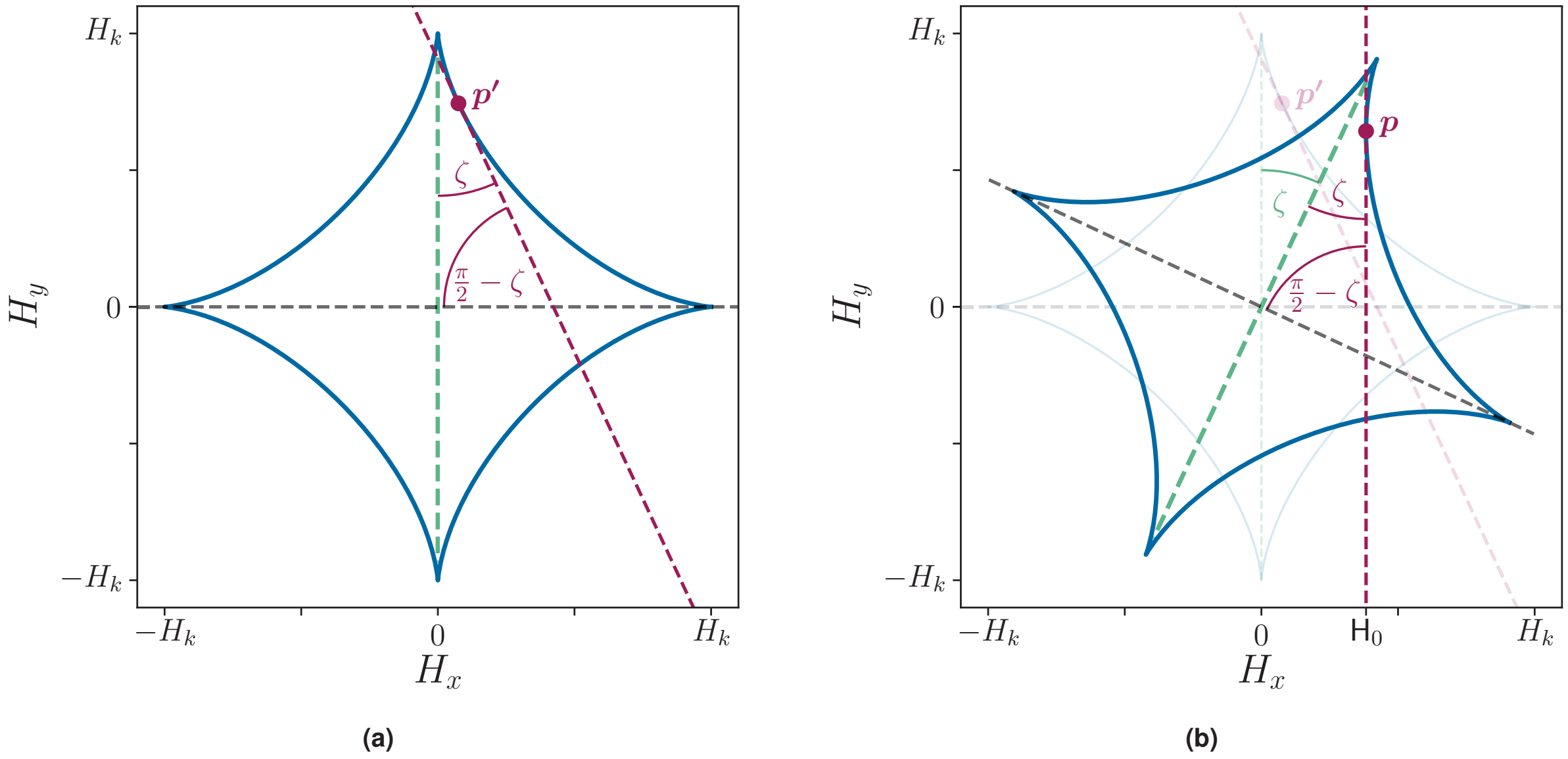

**Figure 6.** Geometric construction of $H_0(\zeta)$ for tilting angle $\zeta$. $H_0$ is found if we draw a tangent to the astroid, parallel to the y-axis. All field vectors drawn from the origin to any point on the tangent, result in one magnetization state with $M_x = 0$ (in this case the down state). It is interesting to note that the zero crossing of $M_x$ does not depend on $H_y$, that a universal zero-magnetization field $H_x = H_0$ can be found and that $H_0$ depends solely on the easy axis tilt angle $\zeta$. The figure shows the construction of $H_0$ for positive $H_x$ and the down state. This choice is completely arbitrary and only motivated by a preference for the first quadrant. The same construction can be done in the other quadrants. The value of the zero-crossing field is given by (19): $H_0 = H_k \sin(\zeta)\cos(\zeta)$

The curves are deformed and separated from each other. Consequently the zero crossings of $M_x$ for up and down state shift to $+H_0$ and $-H_0$ respectively. In the main manuscript is is discussed that for a particular field value tangent to the SW - astroid give the direction of the minimum energy states. For an untilted SW - astroid the only tangent that lead to $M_x = 0$ have to pass the origin of the coordinate system. Hence, $H_0 = H_x = 0$ has to be fullfilled for $M_x = 0$. In other words $H_0$ is only a condition for the x-component of the field, and is furthermore completely independent from the y-component of the external field. The y-component of the external field is merely assumed to be in a region where the state, for which we are looking for a zero-crossing, exists to begin with, i.e. where we can draw a tangent like in fig. 6a. To obtain a value for $H_0$, we need to find the tangential point of the tilted astroid, in the untilted cartesian coordinate system. In parametric form, the untilted astroid can be described by

$$\boldsymbol{r}' = \begin{pmatrix} H_x \\ H_y \end{pmatrix} = H_k \begin{pmatrix} \cos^3 t \\ \sin^3 t \end{pmatrix} \qquad t \in [0, 2\pi] \tag{13}$$

and that the tangent we are looking for (fig. 6a), has a slope of

$$k = -\tan\left(\frac{\pi}{2} - \zeta\right) \tag{14}$$

The tangential point $\boldsymbol{p'}$ on the astroid is found if we equate the first derivative of the astroid boundary with the slope of the tangent.

$$\frac{\mathrm{d}r'_y}{\mathrm{d}r'_x} = \frac{\frac{\mathrm{d}r'_y}{\mathrm{d}t}}{\frac{\mathrm{d}r'_x}{\mathrm{d}t}} = -\frac{\sin^2 t \cos t}{\cos^2 t \sin t} = -\tan t \tag{15}$$

$$-\tan t_{p'} = -\tan\left(\frac{\pi}{2} - \zeta\right) \tag{16}$$

$$t_{p'} = \frac{\pi}{2} - \zeta \tag{17}$$

in the untilted, primed system, the tangential point is therefore given by

$$\boldsymbol{p}' = \boldsymbol{r}'(t_{p'}) = H_\mathrm{k}\begin{pmatrix}\cos^3\left(\frac{\pi}{2}-\zeta\right)\\ \sin^3\left(\frac{\pi}{2}-\zeta\right)\end{pmatrix} = H_\mathrm{k}\begin{pmatrix}\sin^3\zeta\\ \cos^3\zeta\end{pmatrix}. \tag{18}$$

Simply rotating the system by $-\zeta$ yields the tangential point on the tilted astroid $\boldsymbol{p}$.

$$\boldsymbol{p} = \boldsymbol{R}_{-\zeta}\boldsymbol{p}' = H_\mathrm{k}\begin{pmatrix}\cos\zeta & \sin\zeta\\ -\sin\zeta & \cos\zeta\end{pmatrix}\begin{pmatrix}\sin^3\zeta\\ \cos^3\zeta\end{pmatrix} =$$

$$= H_\mathrm{k}\begin{pmatrix}\sin^3\zeta\cos\zeta + \cos^3\zeta\sin\zeta\\ -\sin^4\zeta + \cos^4\zeta\end{pmatrix} = H_\mathrm{k}\begin{pmatrix}\sin\zeta\cos\zeta\\ \cos 2\zeta\end{pmatrix}$$

Since, by definition, the tangent at the unprimed point $\boldsymbol{p}$ is now parallel to the y-axis, the field $H_0$, where the magnetization states have zero crossings, is directly given by the x-component of $\boldsymbol{p}$.

$$H_0 = H_\mathrm{k}\sin\zeta\cos\zeta \tag{19}$$

The sign of $H_0$ depends on the tilting direction and whether the particle is in up or down state (Table 1).

| | $\zeta > 0$ | $\zeta < 0$ |
|---|---|---|
| **up** | $+H_0$ | $-H_0$ |
| **down** | $-H_0$ | $+H_0$ |

**Table 1.** Sign of $H_0$, depending on magnetization state and tilting direction.

## Fundamental phase noise in a Stoner-Wohlfarth model

### Free layer perpendicular to the pinned layer

In order to elucidate the origin of phase noise in SW-particles we discuss the response of a SW particle to rotating fields, where the easy axis of the free layer is perfectly aligned parallel to the y-direction and the pinned layer magnetization is parallel to the x-direction, as indicated in fig.7 (a). In this configuration, the sensor output is sensitive to the average magnetization direction in the x- direction Mx. Exploiting the Stoner-Wohlfarth theory, the free layer is assumed to have homogeneous magnetization. Exposed to an external rotation field Hext, as defined in (1) of the main manuscript, the free layer can only irreversibly switch from up-state ($M_\mathrm{y} > 0$) to down-state ($M_\mathrm{y} < 0$) when the rotating field intersects the Stoner-Wohlfarth (SW) astroid[4] illustrated in fig.7 (a). Conveniently, the SW astroid (blue) can be defined as the envelope of ellipses (red) of the rotating field under the constraint

$$A_\mathrm{x} + A_\mathrm{y} = const = \mu_0 H_\mathrm{k} \tag{20}$$

with $H_\mathrm{k}$ as the anisotropy field. This gives us a theoretical relation for amplitudes of elliptical fields that are tangent to the SW astroid and is thus barely sufficient to allow for the switching of states at $T = 0$. Since in the following phase noise plots, we use $A_\mathrm{y}/A_\mathrm{x}$ rather than $A_\mathrm{y}$ as the abscissa, we rearrange (20): and obtain

$$\frac{A_\mathrm{y}}{A_\mathrm{x}} = \frac{\mu_0 H_\mathrm{k}}{A_\mathrm{x}} - 1 \tag{21}$$

as the condition for fields tangent to the SW astroid, where $A_\mathrm{x}$ and $A_\mathrm{y}$ are the semi-axes of ellipses, which are oriented parallel and perpendicular to the easy axis, respectively. Let us first discuss the free layer response if small fields are applied. The field is sufficiently small so that it does not intersect the SW astroid. For the applied rotating field for a particular $H_\mathrm{x}$ value, there are two opposing Hy components during one rotation cycle. As a consequence for one $H_\mathrm{x}$ component, there are two different equilibrium $M_\mathrm{x}$ configurations, depending on whether the $H_\mathrm{y}$ component is parallel or antiparallel to $M_\mathrm{y}$. This leads to the $M_\mathrm{x}(H_\mathrm{x})$ curve as an open loop with a zero crossing at $H_\mathrm{x} = 0$, as shown in fig.8 (a). The field where $M_\mathrm{x} = 0$ is independent of the initial state and remains zero for the up and down state. In fig.8 (b), the $M_\mathrm{x}$ component is shown as a function of time for an initial up (cyan) and down magnetization (red). Due to different $H_\mathrm{y}$ components over time $t$, the two curves do not coincide. However, the zero crossings $M_\mathrm{x} = 0$ are independent of the state, as already mentioned before. Hence, no matter whether the state points up or down, or even reverses due to thermal fluctuation from one state to the other, the zero crossings of the signal remain constant, and no phase noise is expected. If larger rotating external fields are applied that intersect with the SW astroid, the initial magnetization becomes irrelevant for the $M_\mathrm{x}(t)$ curve, since the field is sufficiently large to switch it in the field direction after the field crosses the SW astroid the first time. As a consequence, the zero crossings are independent of the state again, and phase noise is not observed.

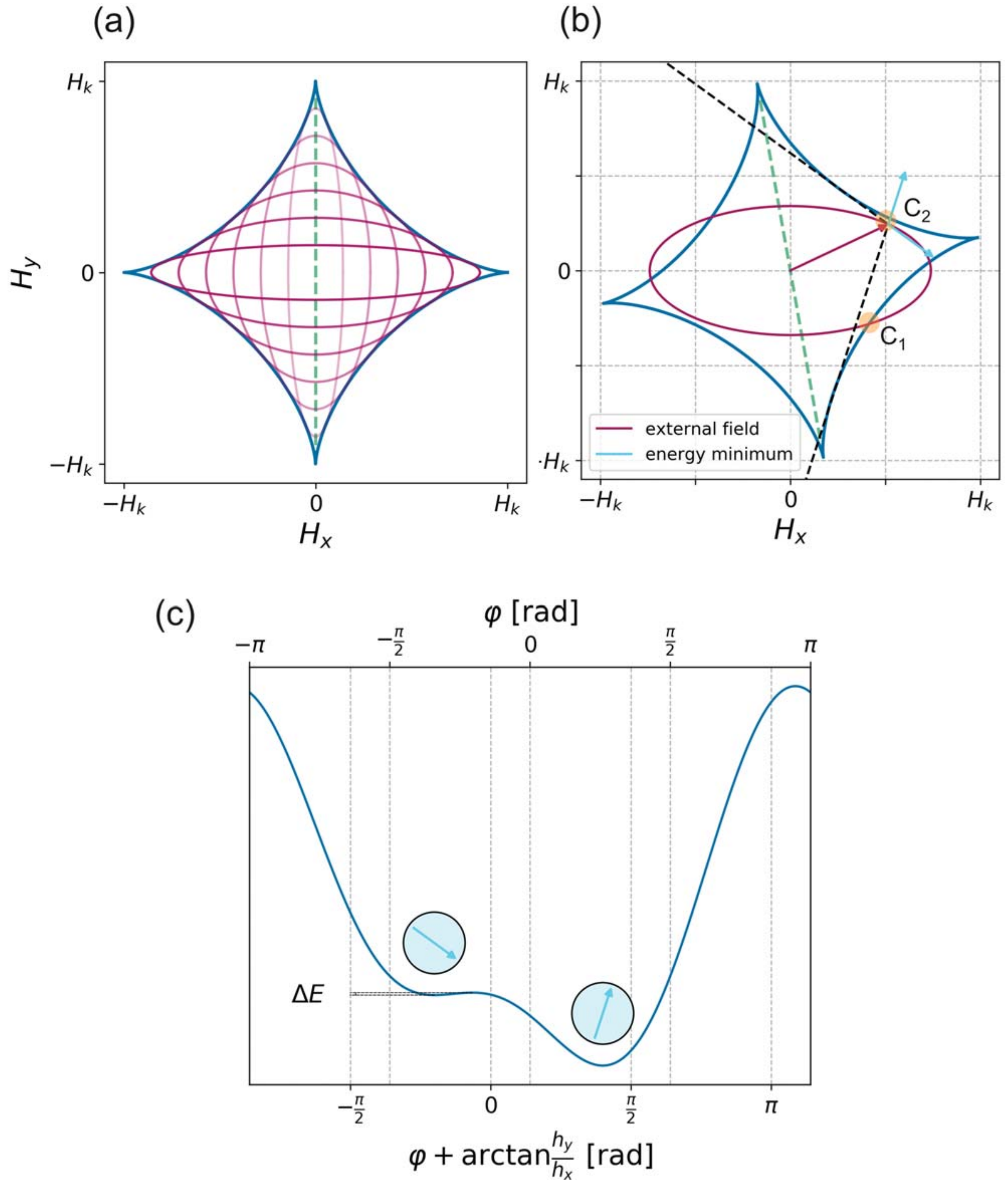

**Figure 7.** (a) Stoner-Wohlfarth astroids with external fields in units of $H_k$. The switching astroid for a Stoner-Wohlfarth particle with an anisotropy field $H_k$ can be constructed as the hull of all ellipses with a constant sum of semi-axes $Ax + Ay$ = const. = $A_k$. (b) A rotating field (red) that is almost tangent to the tilted Stoner-Wohlfarth astroid ($A_x$ = 7.4 mT, $A_y$ = 3.4mT, 0 $H_k$ = 10 mT) in quadrant I is applied for tilting angle $\zeta = 10°$ The equilibrium magnetization of an initially up and down state is shown by the cyan colored arrows at C2. The field is shown by the red arrow. The equilibrium position of the magnetization is found if tangents to the astroid are drawn. The energy of the two equilibrium positions are shown in (c). The angle is the angle between external field and magnetization. Since the field vector is close to the tangent the energy barrier separating the up and down state is small.

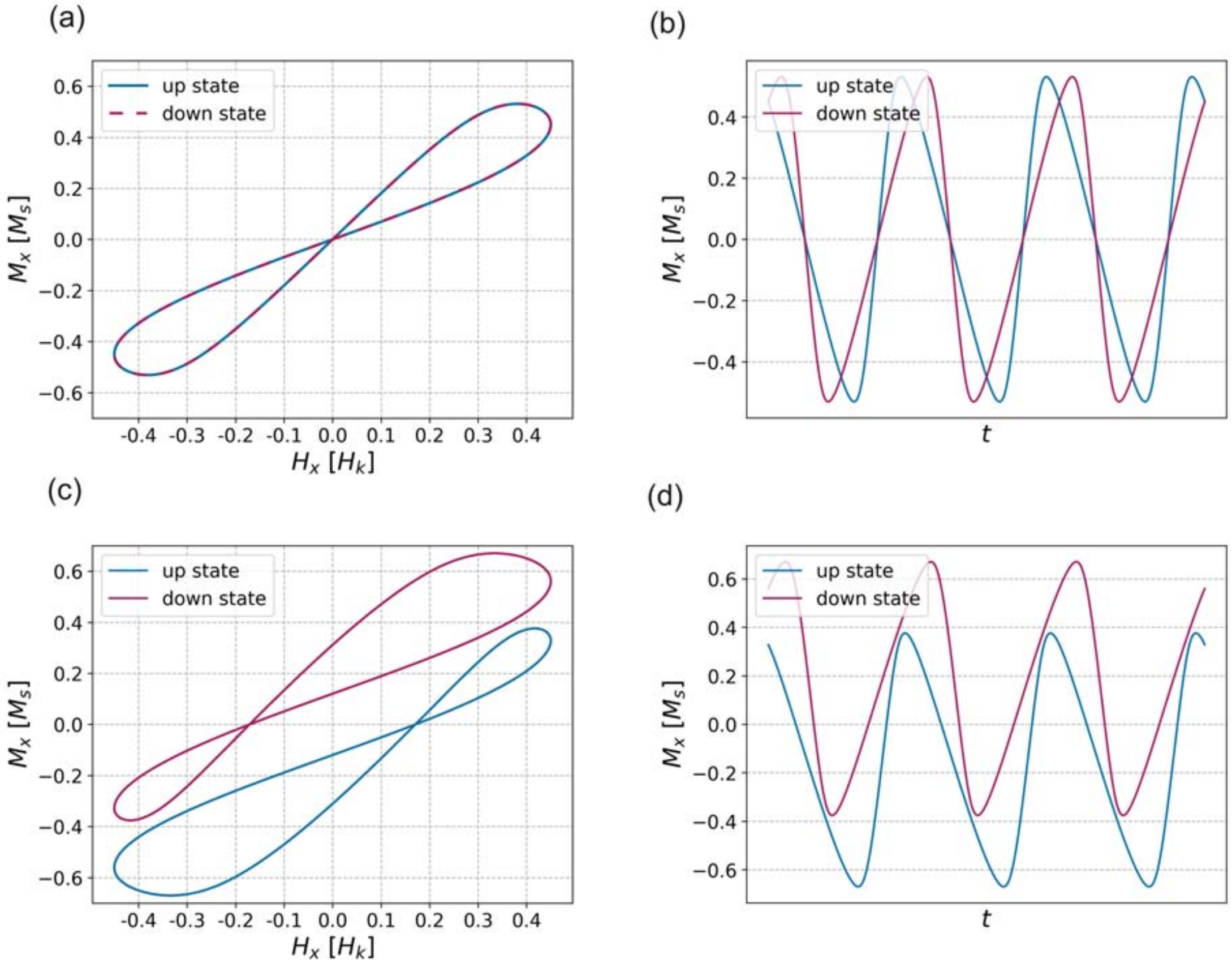

**Figure 8.** Response of the $M_x$ component of the magnetization in the free layer if a rotating external field is applied with $A_x$ = $A_y$ = 4.5 mT and 0 $H_k$ = 10 mT. For the cyan and red curves, the particle is initialized in the up and down directions, respectively. For (a) and (b), the angle between the y-axis and the easy axis is $\zeta = 0°$. For (c) and (d), $\zeta$ = 10°. (a) and (c) show $M_x$ as a function of one component ($H_x$) of the applied field. (b) and (d) show the time evolution of $M_x$ and the rotating field.

### Free layer not perpendicular to the pinned layer

The situation becomes different if the easy axis of the free layer is no longer exactly perpendicular to the pinned layer, which is a realistic case due to alignment inaccuracy in the fabrication or annealing step. Let us assume that the pinned layer is still oriented in the x-direction and discuss the situation, where the applied fields do not intersect the SW astroid. The $M_x$ component of the magnetization as a function of $H_x$ is shown in fig.8 (c) for the case of an initial up state (cyan) and down state (red), respectively. It can be seen that the zero crossing of $M_x$ depends on whether the free layer is initially in the up or down state. This behavior is also illustrated in fig. 8 (d), where $M_x(t)$ is shown. Since we restrict the discussion here to small fields, the free layer does not switch the state. This means that the zero crossing of the $M_x$ is completely reproducible and leads to zero phase noise. However, for given field amplitudes that are almost tangent to the SW astroid, as shown in fig.7 (b), the situation becomes different. Let us assume that the external field (red) rotates counterclockwise in Fig. 3 (c). The tilting angle of $\zeta = 10°$ is arbitrarily chosen to visualize the phenomenon. A tilting angle up to a few degrees is plausible in reality. For fields that intersect with the tilted SW astroid (blue), the particle will switch states when the field crosses the SW astroid boundary, beyond which the current particle state is no longer an energy minimum. The outside regions of quadrants I and II correspond to the up state, while quadrants III and IV correspond to the down state. Hence, when the field intersects the SW astroid in point C1, the particle will be set to the down state. When the field gets close to the SW astroid, as in state (C2), the particle might switch to the up state or may remain in the down state due to the stochastics of finite temperature[5–7]. A detailed description of how finite temperature is included in the simulations by a kinetic Monte Carlo approach is given in the next section. The switching probability depends on temperature, the rotation frequency, magnetic parameters, and the detailed field strength. Depending on whether the free layer switches into the up state or remains in the down state, the response to the rotating field is different, as discussed previously. Consequently, the zero crossing of $M_x(H_x)$ happens at different external fields $H_x$ and, hence, at different times. The field $H_x$ where the zero crossing occurs is given by (19), where $\zeta$ is the angle between the y-axis and the free layer easy axis. The easy axis tilt angle is measured from the positive y-axis in such a manner that a positive tilting angle leads to an easy axis that lies in the second and fourth quadrants of the Cartesian coordinate system. The sign of $H_x$ depends on whether the particle is in the up or down state. While (19) holds true for the up state, a negative sign has to be introduced for the down state.

## Relation between detectivity $D$ and minimum detectable field amplitudes

Let us assume that we want to detect a sinusoidal magnetic field $B_z$ that oscillates with a fixed and known angular frequency $\omega_i$. We aim to estimate the minimum amplitude $B_0$ that can be detected within a certain measuring time $T$ .

$$B_z(t) = B_0 \cos(\omega_i t) \tag{22}$$

Let us assume we detect the magnetic field with a field sensor that has in the considered linear range of the sensor a constant sensitivity $\mathrm{d}V/\mu_0 \mathrm{d}H_{ext}$. Then the induced voltage due to the magnetic field is given by,

$$u(t) = B_0 \frac{\mathrm{d}V}{\mu_0 \mathrm{d}H_{ext}} \cos(\omega_i t) = u_0 \cos(\omega_i t) \tag{23}$$

In addition to the signal, a noise $n(t)$ is assumed, leading to the total signal

$$u_{tot}(t) = u_0 \cos(\omega_i t) + n(t) \tag{24}$$

In the following we assume that the noise $n(t)$ is white noise with the properties

$$\langle n(t) n(t+\tau) \rangle = \sigma_n^2 \delta(\tau) \tag{25}$$

$\sigma_n^2$ is connected with the spectral density, that is defined as

$$S_V = \frac{1}{2\pi} \int_{-\infty}^{\infty} \langle n(t) n(t+\tau) \rangle e^{-i\omega\tau} d\tau \tag{26}$$

Which leads for white noise to the relation

$$S_V = \frac{\sigma_n^2}{2\pi} \tag{27}$$

Let us assume we measure the total signal with a lock – in amplifier, as an example of the most narrow band path filter as possible. Then the output due to the signal is given by

$$S_{\text{signal}} = \frac{1}{T}\int_0^T u(t)\cos(\omega_\text{i} t)\,\mathrm{d}t = \frac{1}{T}\int_0^T u_0 \cos^2(\omega_\text{i} t)\,\mathrm{d}t = \frac{1}{2}u_0 \tag{28}$$

The output of the lock – in amplifier due the noise is

$$S_{\text{noise}} = \frac{1}{T}\int_0^T n(t)\cos(\omega_\text{i} t)\mathrm{d}t \tag{29}$$

In order to resolve the output $S_{\text{noise}}$ from the noise, it is required that the standard deviation of the noise output $\sigma(S_{\text{noise}})$ is smaller than the signal. The lowest amplitude $u_0$ that can be distinguished from zero within the confidence limit of 1 % is given by[8],

$$3\sigma(S_{\text{noise}}) < S_{\text{signal}} \tag{30}$$

In order to calculate $\sigma(S_{\text{noise}})$ we introduce a measuring time that is a multiple of the signal period,

$$T = k\tau_{\text{sig}} \tag{31}$$

with

$$\omega_\text{i} = \frac{2\pi}{\tau_{\text{sig}}} \tag{32}$$

leads to

$$S_{\text{noise}} = \frac{1}{T}\int_0^T n(t)\cos\left(\frac{2\pi}{\tau_{\text{sig}}}t\right)\mathrm{d}t = \frac{1}{k\tau_{\text{sig}}}\int_0^{k\tau_{\text{sig}}} n(t)\cos\left(\frac{2\pi}{\tau_{\text{sig}}}t\right)\mathrm{d}t \tag{33}$$

In order to calculate the standard deviation of the random variable $S_{\text{noise}}$, we rewrite (29) with a Wiener Process

$$\mathrm{d}S_{\text{noise}} = \frac{1}{k\tau_{\text{sig}}}\cos\left(\frac{2\pi}{\tau_{\text{sig}}}t\right)\sigma_\text{n}\mathrm{d}W \tag{34}$$

with variance $\sigma_\text{n}^2$ . We obtain for $t = k\tau_{\text{sig}}$

$$\sigma(S_{\text{noise}}) = \frac{\sigma_\text{n}}{\sqrt{2k\tau_{\text{sig}}}} = \sqrt{\frac{\pi S_\text{V}}{k\tau_{\text{sig}}}} \tag{35}$$

Hence, the minimal signal amplitude $u_0$ that can be detected within a confidence limit of 1% is given by,

$$3\sqrt{\frac{\pi S_\text{V}}{k\tau_{\text{sig}}}} < \frac{1}{2}u_0 \tag{36}$$

which can be written with the detectivity $D$ as:

$$6D\sqrt{\frac{\pi}{T}} \approx \frac{10D}{\sqrt{T}} < B_0 \tag{37}$$

## Additional information

The corresponding author is responsible for submitting a competing financial interests statement on behalf of all authors of the paper.